%
\magnification=1200
\tolerance=1400
\overfullrule=0pt
\baselineskip=11pt

\font\rmc=cmr9
\font\rmb=cmr9 scaled \magstep 1
\font\rma=cmr9 scaled \magstep 2
\font\rmm=cmr9 scaled \magstep 3
\font\rmw=cmbx9  scaled \magstep 2
 2
\font\tenib=cmmib10
\font\sevenib=cmmib10 at 7pt
\font\fiveib=cmmib10 at 5pt
\newfam\mitbfam
\textfont\mitbfam=\tenib
\scriptfont\mitbfam=\sevenib
\scriptscriptfont\mitbfam=\fiveib

\mathchardef\bfo="0\the\mitbfam21
\mathchardef\om"0\the\bffam0A
 3
\font\bigastfont=cmr10 scaled \magstep 3

\input mssymb
\def\ueber#1#2{{\setbox0=\hbox{$#1$}%
  \setbox1=\hbox to\wd0{\hss$\scriptscriptstyle #2$\hss}%
  \offinterlineskip
  \vbox{\box1\kern0.4mm\box0}}{}}

\def\R{\rm I\kern-.18em R}
\def\ref{\par\noindent\hangindent\parindent\hangafter1}
\def\etal{{\it et al. }}
\def\1{_{\vert}}
\def\bs{\lower6pt\hbox{\bigastfont *}}
\font\bigastfont=cmr10 scaled \magstep 3

\topskip 6 true cm
\pageno=0

\centerline{\rmw NEWTONIAN COSMOLOGY}
\medskip
\centerline{\rmw IN LAGRANGIAN FORMULATION:}
\bigskip
\centerline{\rmw FOUNDATIONS AND PERTURBATION THEORY}
\bigskip\bigskip
\centerline{\rmw by}
\bigskip\bigskip
\centerline{\rmw J\"urgen Ehlers$^1$ \& Thomas Buchert$^2$}
\vskip 1.0 true cm
\centerline{\bf $^1$ Max--Planck--Institut f\"ur Gravitationsphysik}
\smallskip
\centerline{\bf Albert--Einstein--Institut}
\smallskip
\centerline{\bf Schlaatzweg 1}
\smallskip
\centerline{\bf D--14473 Potsdam}
\smallskip
\centerline{\bf F. R. G.}
\vskip 0.7 true cm
\centerline{\bf $^2$ Ludwig--Maximilians--Universit\"at}
\smallskip
\centerline{\bf Theoretische Physik}
\smallskip
\centerline{\bf Theresienstr. 37}
\smallskip
\centerline{\bf D--80333 M\"unchen}
\smallskip
\centerline{\bf F. R. G.}
\vskip 2 true cm
\centerline{\bf submitted to {\bf G.R.G.}}
\medskip
\centerline{\bf Version of September 3$^{rd}$, 1996}
\vfill
\eject

\topskip 3 true cm
\pageno=0
\centerline{\rmm Newtonian Cosmology in Lagrangian Formulation:}
\smallskip
\centerline{\rmm Foundations and Perturbation Theory}
\bigskip\bigskip
\centerline{\rma by}
\bigskip\bigskip
\centerline{\rma J\"urgen Ehlers$^1$ \& Thomas Buchert$^2$}
\vskip 1.0 true cm
\noindent
{\rmc
{\narrower
{\rma Summary:}
The ``Newtonian'' theory of spatially unbounded, self--gravitating,
pressureless continua in Lagrangian form is reconsidered. 
Following a review of the pertinent kinematics, we present alternative 
formulations of the Lagrangian evolution equations and establish conditions
for the equivalence of the Lagrangian and Eulerian representations. We then
distinguish open models based on Euclidean space $\R^3$ from closed models
based (without loss of generality) on a flat torus ${\Bbb T}^3$.
Using a simple averaging method we show that the spatially averaged variables
of an inhomogeneous toroidal model form a spatially homogeneous ``background''
model and that the averages of open models, if they exist at all, in general
do not obey the dynamical laws of homogeneous models. We then specialize to 
those inhomogeneous toroidal models whose (unique) backgrounds have a 
Hubble flow, and derive Lagrangian evolution equations which govern the
(conformally rescaled) displacement of the inhomogeneous flow with respect
to its homogeneous background. Finally, we set up an iteration scheme and
prove that the resulting equations have unique solutions at any order for
given initial data, while for open models there exist infinitely many
different solutions for given data.

}}

\bigskip\bigskip\bigskip\noindent
$^1$ e--mail: ehlers@aei--potsdam.mpg.de
\medskip\noindent
$^2$ e--mail: buchert@stat.physik.uni--muenchen.de

\vfill\eject
\topskip= 0 true cm
\pageno=1
\noindent{\rmm 1. Introduction}
\bigskip\bigskip
\noindent
The Lagrangian theory of gravitational instability of
Friedmann--Lema\^\i tre cosmologies
turned out to be a much more powerful tool for the modeling of
inhomogeneities in Newtonian cosmology than the standard
Eulerian perturbation approach was
(for the latter see, e.g., Peebles 1980, 1993, and ref. therein).
\medskip
Already the general first--order solution of this
theory (Buchert 1989, 1992)
(which contains the widely applied ``Zel'dovich approximation",
Zel'dovich 1970, 1973, as a special case)
has been found to provide an excellent approximation
of the density field in the weakly non--linear regime (i.e., where the
r.m.s. deviations of the
Eulerian density contrast field
$\delta:= \rho / \rho_H -1$ are of order
unity) in contrast to the Eulerian linear theory of gravitational
instability (Coles \etal 1993, Buchert \etal 1994, Bouchet \etal 1995,
Sahni \& Coles 1995).
This appears to be due to the fact that, in contrast to the Eulerian scheme,
the Lagrangian approximation takes fully into account, at any order, the 
convective part $(\vec v \cdot \nabla ) \vec v$ 
of the acceleration and conservation of mass. Another advantage of the
Lagrangian equations is that they are regular at caustics (where the 
density blows up), whereas Euler's equations break down. Therefore, 
Lagrangian solutions can be continued accross caustics, i.e., at the places
where structures form.
\medskip
Most recently, the range of application of
Lagrangian perturbation solutions for the modeling of large--scale
structure has been greatly extended by employing filtering techniques
which discard high--frequency modes in the power--spectrum of the
initial data, and so enable to model highly non--linear stages, even
in hierarchical models with much small--scale power
(Melott \etal 1994, 1995, Wei{\ss} \etal 1996).

\medskip
In view of these results we think that the power of the
Lagrangian description, usually applied only to flows under very
restrictive conditions (planar, incompressible, etc.), has
been underestimated.
The recent investigation of solutions demonstrates
that the complicated nonlinear partial differential equations which
result from the transformation of the Eulerian equations to
Lagrangian coordinates {\it can be} solved in special cases even 
in three dimensions (see Subsection 3.2.3), which was claimed to be
impossible in standard text books on hydrodynamics discussing the
Lagrangian
picture. One reason for the possibility of constructing solutions lies
in the close correspondence of Lagrangian flows and classical point
mechanics: the Lagrangian coordinates {\it label} fluid elements
like coordinate indices, and in perturbation theory the Lagrangian evolution
equations for dust reduce to a sequence of {\it ordinary} 
differential equations, as will be shown below.
\smallskip
For details on the Lagrangian picture of fluid motion in classical
hydrodynamics see Serrin (1959) and the compilation by Stuart \& Tabor (1990).
\medskip
We shall treat the initial value problem for the Lagrangian perturbation 
equations of all orders, using a global gauge condition to fix the relation
between the background and the perturbed flows, and we establish existence
and uniqueness of perturbative solutions for toroidal (or spatially periodic)
models, thus complementing work by Brauer (1992) and Brauer \etal (1994).
\medskip
Lagrangian perturbation theory has become popular; various authors
pursue similar studies
in relation to the modeling of large--scale structure
in the Universe (Moutarde \etal 1991, Bouchet \etal 1992,
Lachi\`eze-Rey 1993, Gramann 1993, Munshi \etal 1994). For reviews
see Bouchet \etal (1995), Bouchet 1996,
Sahni \& Coles (1995) and Buchert (1996a,b).
\medskip

Recent efforts concerning the Lagrangian theory in general relativity and 
in particular Langrangian perturbation solutions have been also focussed 
on evolution equations for fluid quantities such as
shear and vorticity, the gravitational tidal tensor as the ``electric part''
of the Weyl--tensor, as well as the ``magnetic part'' of the Weyl--tensor.
Supported by the classical works by Ehlers (1961), Tr\"umper (1965) and
Ellis (1971), a variety of perspectives in cosmology have been opened, see
the works by Kasai (1992, 1995), Matarrese \etal (1993, 1994),
Croudace \etal (1994), Salopek \etal (1994), Bertschinger \& Jain (1994),
Bertschinger \& Hamilton (1994), Bruni \etal (1995), Kofman \& Pogosyan (1995), 
Lesame \etal (1996), Ellis \& Dunsby (1996), Bertschinger (1996),
Matarrese (1996), Matarrese \& Terranova (1996), Russ \etal (1996).
In these works also the Newtonian limits, or analogues, respectively,
have been discussed. In a separate note we
complement this focus by giving a clear--cut definition of the
Newtonian limits of the electric and magnetic parts of the Weyl--tensor
in a $4-$dimensional ``frame theory'' which covers both Newton's and
Einstein's theory (Ehlers \& Buchert 1996). In Newton's theory such
fluid quantities are expressed in terms of functionals of the trajectories. 
We emphasize that our point of view of a Lagrangian treatment of 
evolution equations,
which was begun with the formulation of a closed Lagrangian system
for the trajectories by Buchert \& G\"otz (1987), aims to determine
fluid quantities explicitly in terms of the trajectory field, and even
integrate these quantities along the trajectories, if possible,
thus, reducing the description to a {\it single} dynamical field variable.
This point of view enables to determine explicitly the evolution of
fluid quantities without specifying particular solutions for the
trajectories. 

\bigskip
\medskip\noindent
The paper is organized as follows:
\smallskip\noindent
In Section 2 we summarize some pertinent facts on the kinematics
and dynamics of Newtonian self--gravitating flows
in the Lagrangian framework. We give an alternative formulation of
the Lagrangian evolution equations in terms of differential forms,
we address the initial value problem, the problem of existence of
solutions, and the equivalence of Eulerian and Lagrangian formulations
up to the stage when shell--crossing singularities occur.
We aim to give a self--contained representation of the equations
and some additional useful relations.
Therefore, some equations are reviewed which are not needed
in the following sections

In Section 3 we discuss the Lagrangian theory of gravitational
instability of the Newtonian analogues of Friedmann cosmologies. Here, we
give the general perturbation and solution schemes at any order and discuss
the modeling of space as a $3-$torus ${\Bbb T}^3$ as compared to
$\R^3$.
We give detailed remarks on the interpretation of the perturbation
scheme and prove uniqueness of the perturbation solutions at any order
on the $3-$torus.

\vfill\eject

\noindent{\rmm 2. The Lagrangian framework}
\bigskip\bigskip
\noindent
{\rma 2.1. Kinematics}
\bigskip\noindent
{\rmb 2.1.1. Integral--curves and displacement maps}
\bigskip\noindent
Let $\vec v \lbrack \vec x,t \rbrack$ denote a smooth Eulerian velocity
field on
${\R}^3 \times \lbrack t_0,t_1 \rbrack$.

\noindent
We assume that $\vert \vec v \vert \le V, \vert
\partial v_i  / \partial x_j \vert \le M$ (indices run from $1$ to $3$)
\footnote*{We employ orthonormal coordinates $x_i$ and use
corresponding vector and tensor components; therefore
all indices may be written as subindices.}.
Then there exists a unique
smooth vector field $\vec f (\vec X,t)$ such that
$$
\eqalignno{& {d \vec f \over dt} = \vec v \lbrack \vec f (\vec X,t), t
\rbrack \;\;,\;\;
\vec f(\vec X, t_0) = :\vec X \;\; . &(1a,b)\cr}
$$
The {\it integral--curves} $t \mapsto \vec x (t) = \vec f (\vec X,t)$ of
the velocity field are labelled
by the (initial) Lagrangian coordinates $\vec X$; $d /dt:= \partial /
\partial t
+ \vec v \cdot \nabla$ is the total
(Lagrangian) time derivative, henceforth abbreviated by a dot;
a comma (or $\nabla$) denotes
differentiation with respect to Eulerian coordinates, and a
vertical slash (or $\nabla_0$)
denotes differentiation with respect to Lagrangian
coordinates; only the latter commutes with the dot. Since dependent
variables will sometimes be expressed either in terms of Eulerian
or in terms of Lagrangian coordinates, we emphasize the different
functional dependence by using the notations $\lbrack \vec x,t \rbrack$
or $(\vec X,t)$, respectively.

\smallskip
Our assumptions on $\vec v$ imply the following statements $(A-G)$:
$$
{\rm The}\;{\rm integral-curves}\;{\rm defined}\;{\rm by}\;\;\vec f
\;\;{\rm do}\;{\rm not}\;{\rm
intersect}\;\;.\eqno(A)
$$
Since the volume expansion rate $\theta : =
\nabla \cdot \vec v$ is bounded by $3M$, and since (1) gives
for the Jacobian
$$
J:=\det (f_{i\1 k})\eqno(2a)
$$
the equation
$$
J(\vec X,t) = e^{\int_{t_0}^{t} dt' \;\theta \lbrack \vec f (\vec
X,t'),t'\rbrack}
\;\;,\eqno(2b)
$$
we obtain
$$
0 < e^{-3M(t_1-t_0)} \le J(\vec X,t) \le e^{3M(t_1-t_0)}\;\;.\eqno(B)
$$
Due to (1a),
$$
\vert {\dot{\vec f}} \vert \le V \;\;.\eqno(C)
$$
The definition (1a,b) of $\vec f$ implies that
$$
{\dot f}_{i\1 k} = v_{i,\ell} f_{\ell \1 k} \;\;;\eqno(2c)
$$
therefore, the elements of the {\it deformation gradient} $\nabla_0 \vec f$
are bounded,
$$
\vert f_{i\1 k} \vert \le e^{3M(t_1 - t_0)} \;\;,\eqno(D)
$$
and
$$
\vert {\dot f}_{i\1 k} \vert \le 3M e^{3M(t_1 - t_0)} \;\;.\eqno(E)
$$
These properties further have the consequences that
the {\it displacement map}
$f_t: \vec X \mapsto \vec x = \vec f (\vec X,t)$, which sends
fluid particles from their initial positions at time $t_0$ to their
positions at time $t$, has the following property:
$$
f_t \;\;{\rm is}\;{\rm an}\;{\rm orientation}\;{\rm preserving}\;{\rm
diffeomorphism}\;{\rm of}\;\;{\R}^3 \;\; {\rm onto}\;{\rm itself}
\;\;,\eqno(F)
$$
(see Appendix A for a proof).

Let $h_t$ denote the inverse of $f_t$;
$\vec X = \vec h \lbrack\vec x,t\rbrack$. Its Jacobian matrix is given
by
$$
h_{j,\ell} = {1\over
2J}\epsilon_{\ell pq}\epsilon_{jrs}f_{p\1 r}f_{q\1 s}\;\;, \eqno(3a)
$$
and therefore
$$
\vert h_{j,\ell} \vert \le e^{9M(t_1 - t_0)} \;\;.\eqno(G)
$$
\medskip
So far, we have listed consequences of the definition (1a,b) of $\vec f$
in terms of $\vec v$. Let us now, conversely, assume that we have a
smooth $\vec f (\vec X,t)$ which has, on $\R^3 \times \lbrack t_0, t_1
\rbrack$, the properties $(A),(B),(C),(D),(E)$. Then it is
easily established that $(F)$ and $(G)$ also hold, and the Eulerian
velocity field
$$
{\vec v} \lbrack \vec x,t \rbrack := {\dot{\vec f}}({\vec h} \lbrack
\vec x,t\rbrack,t) \eqno(3b)
$$
is smooth and enjoys boundedness properties of the kind we started with.
These remarks show under which assumptions the kinematics defined
by an Eulerian $\vec v \lbrack x,t \rbrack$ or a Lagrangian $\vec f
(\vec X,t)$, respectively, are equivalent; we then call the kinematics
{\it regular}.

\bigskip\noindent
{\bf Remarks:}
\smallskip\noindent
(i) The preceding statements remain valid, with some adaptations,
if space is modeled not as $\R^3$, but as a torus ${\Bbb
T}^3$.
\smallskip\noindent
(ii) If, contrary to our
assumptions, the velocity field $\vec v$ or $\dot{\vec f}$ were
not bounded, fluid particles might escape to infinity in a finite time.
If $\theta \rightarrow -\infty$
sufficiently fast, then $J\rightarrow 0$ there, and $f_t$ would
no longer be locally diffeomorphic;
the flow would then develop a caustic. If (A) were violated,
$f_t$ would no longer be injective. In all three cases, (F) would fail.

\bigskip\noindent
Under the assumptions discussed above we can also obtain the
Eulerian {\it acceleration field} $\vec g = {\vec v}_{,t} + \vec v \cdot
\nabla \vec v$ from $\vec f$:
$$
{\vec g} \lbrack \vec x,t \rbrack := {\ddot{\vec f}}({\vec h}
\lbrack\vec x,t\rbrack,t)
\;\;.\eqno(3c)
$$
It is convenient to introduce the following abbreviation:
Calculating the Eulerian velocity gradient we obtain, with (3a),
$$
v_{i,\ell} = {\dot f}_{i\1 j}h_{j,\ell} = {1\over 2} \epsilon_{\ell pq}
{\cal J} ({\dot f}_i,f_p,f_q)J^{-1}
\;\;,\eqno(3d)
$$
where ${\cal J}(A,B,C)$ abbreviates the
functional determinant of any three functions $A(\vec X,t)$,
$B (\vec X,t)$, $C(\vec X,t)$ with respect to Lagrangian
coordinates:
$$
{\partial (A, B, C)\over
\partial(X_1,X_2,X_3)}=:{\cal J}(A,B,C)\;\;,
$$
e.g., for the Jacobian we simply have $J={\cal J}(f_1,f_2,f_3)$.

\bigskip
\noindent
We now write the curl and the divergence of $\vec g$ in terms of
$\vec f$, using $\vec h$ as a transformation from
Eulerian to Lagrangian coordinates (hereafter, repeated indices imply
summation, with $i,j,k$ running through the cyclic permutations of
$1,2,3$):
$$
\eqalignno{
(\nabla \times \vec g)_k &=
\epsilon_{pq \lbrack j} {\cal J}({\ddot f}_{i
\rbrack},f_p,f_q) J^{-1}
\;\;,&(4a,b,c)\cr
(\nabla \cdot \vec g) &=
{1 \over 2}\epsilon_{abc}\;{\cal J}({\ddot f}_a,f_b,f_c) J^{-1}
\;\;.&(4d)\cr}
$$
Explicitly, these equations read (summation over j !):
$$
(\nabla \times \vec g)_i =
{\cal J}({\ddot f}_j, f_i, f_j)\; J^{-1}\;\;,
\eqno(4a,b,c)
$$
$$
(\nabla \cdot \vec g) =
\left({\cal J}({\ddot f}_1, f_2, f_3)+{\cal J}({\ddot f}_2,f_3,f_1)
+{\cal J}({\ddot f}_3,f_1,f_2)\right) J^{-1} \;\;.\eqno(4d)
$$
The arguments on the left are $\vec x,t$, on the
right, $\vec h\lbrack\vec x,t\rbrack,t$.
\medskip
Below we give an alternative formulation by using differential forms:

\noindent
Let $d$ denote the operator of {\it spatial} exterior differentiation acting
on functions and forms which may be expressed for regular kinematics
either in
Eulerian $(\vec x)$ or Lagrangian $(\vec X)$ coordinates.
Then, equations (4) read:
$$
{1\over 2}
(\nabla \times \vec g)_i \; \epsilon_{ijk}
dx_j \wedge dx_k \;=\; g_{\lbrack i,j \rbrack} dx_j \wedge dx_i \;=\;
d{\ddot f}_i \wedge df_i  \;=\; d({\ddot f}_i df_i)\;\;,
\eqno(4a,b,c)
$$
and
$$
(\nabla \cdot \vec g) \; dx_1 \wedge dx_2 \wedge dx_3
\;=\;3 d{\ddot f}_{\lbrack 1} \wedge df_2 \wedge
df_{3\rbrack} \;=\;
d(*{\ddot f}_i df_i)\;\;,
\eqno(4d)
$$
where here $*$ denotes the Hodge star operator with respect to the
Euclidean metric $\ueber{dx}{\rightarrow}^2$.
We shall, however, work
with the first form of equation (4d) which turns out to be more
convenient than the more elegant second form. Also, we shall later use
the Hodge star operator with respect to the metric
$\ueber{dX}{\rightarrow}^2$
which coincides with the Euclidean metric
$\ueber{dx}{\rightarrow}^2$ only at $t=t_0$. The latter operator we
shall denote with $\bs$.

Recall that the anti--symmetric part taken over $3$ indices multiplied
by $3$ coincides with the sum of all cyclic permutations in
expressions which involve wedge products, e.g.,
$$
3 d{\ddot f}_{\lbrack 1} \wedge df_2 \wedge df_{3\rbrack} \;=\;
d{\ddot f}_1 \wedge df_2 \wedge df_3 +
d{\ddot f}_2 \wedge df_3 \wedge df_1 +
d{\ddot f}_3 \wedge df_1 \wedge df_2 \;\;.
$$

\bigskip\bigskip
\noindent
{\rmb 2.1.2. Principal invariants of a linear map}
\bigskip\noindent
A linear map ${\cal A}:{\R}^3 \rightarrow
{\R}^3$ has the following three principal scalar invariants:
$$
\eqalignno{
I({\cal A}) :&= tr({\cal A}) \;\;,&(5a)\cr
II({\cal A}):&={1 \over 2} \left( (tr({\cal A})^2 -
tr({\cal A}^2)\right) \;\;, &(5b)\cr
III({\cal A}):&= \det({\cal A})
\;\;. &(5c)\cr}
$$
For cartesian components, ${\cal A}=(A^{i}_j)=(A_{ij})$.

In previous work
the symbols $I$, $II$, and $III$
for the principal invariants of
any linear map have been used,
either with respect to Eulerian or Lagrangian coordinates.
The kinematical scalars for the expansion, the shear, and the vorticity
of the flow $\vec f (\vec X,t)$, which we shall use in the present work,
can be expressed in terms of the
principal invariants (5), which we shall do now.

\bigskip\bigskip
\noindent
{\rmb 2.1.3. Relation to kinematical variables}
\bigskip\noindent
Let us split the Eulerian velocity gradient $(v_{i,j})$
into its symmetric and
anti--symmetric parts,
$$
v_{i,j}=v_{(i,j)}+v_{\lbrack i,j
\rbrack}=:\theta_{ij}+\omega_{ij} \;\;,\eqno(6a)
$$
the {\it rate of deformation} $\theta_{ij}$ and the {\it rate of
rotation} $\omega_{ij}$.
We can split $\theta_{ij}$
into its trace--free part, the (symmetric) {\it shear tensor}
$\sigma_{ij}$, and its trace $\theta$, which was introduced already,
$$
\theta_{ij} = \sigma_{ij} + {1\over 3}\delta_{ij}\theta \;\;.\eqno(6b)
$$
The (anti--symmetric) tensor $-\omega_{ij}$ is dual to the
{\it angular velocity} ${\vec \omega}$, defined as
$$
{\vec \omega}:= {1\over 2} \nabla \times \vec v
\;\;.\eqno(6c)
$$
The vorticity tensor $\omega_{ij} = -\epsilon_{ijk} \omega_k$
can be expressed in terms of $\vec f$,
$$
\omega_{ij} = v_{\lbrack i,j\rbrack}\;=\;{1\over 2}
\epsilon_{pq\lbrack
j} \,{\cal J}({\dot f}_{i\rbrack},f_p,f_q) J^{-1}\;\;,\eqno(6d)
$$
or, using differential forms,
$$
{\bfo}:=
-\omega_{ij} dx_i \wedge dx_j = d{\bf v} = d(v_j dx_j) =
d{\dot f}_j \wedge df_j \;\;.\eqno(6e)
$$
The components of $\vec \omega$, $\omega_i = -
{1\over 2}\epsilon_{ijk}\omega_{jk}$ can be written explicitly as
(summation over j !)
$$
\omega_i = {1\over 2} {\cal J}({\dot f}_j, f_i,
f_j) J^{-1}\;\;.\eqno(6f)
$$
The magnitudes of shear and rotation are given by
$$
\sigma:= \sqrt{{1\over 2}\sigma_{ij}\sigma_{ij}}\;\;\;;\;\;\;
\omega: = \vert \vec \omega \vert =
\sqrt{{1\over 2}\omega_{ij}\omega_{ij}}\;\;.\eqno(6g,h)
$$
The preceeding definitions imply
$$
{1\over 2}\;v_{i,j}v_{i,j}\;=\;
\omega^2 + \sigma^2 + {1\over 6}\theta^2
\;\;,\eqno(7a)
$$
$$
{1\over 2}\;v_{i,j}v_{j,i}\;=\;
-\omega^2 + \sigma^2 + {1\over 6}\theta^2
\;\;.\eqno(7b)
$$
\medskip
In view of (6) and (7)
the principal scalar invariants $I$, $II$ and $III$ of the tensor
$(v_{i,j})$ are expressible in terms of kinematical scalars,
$$
\eqalignno{
I(v_{i,j}) &= v_{i,i} = \nabla \cdot \vec v = \theta \;\;,&(8a)\cr
II(v_{i,j})&= {1\over 2}\left((v_{i,i})^2 - v_{i,j}v_{j,i}\right) 
={1\over 2}\nabla\cdot
\left(\vec v \nabla\cdot\vec v - \vec v \cdot \nabla \vec v \right)
= \omega^2 - \sigma^2 + {1 \over 3} \theta^2
\;\;,&(8b)\cr
III(v_{i,j})&= {1\over 3} v_{i,j}v_{j,k}v_{k,i} - {1\over 2}(v_{i,i})
(v_{i,j}v_{j,i})+{1\over 6} (v_{i,i})^3
={1\over 3}\left( v_i {\cal V}_{ij}\right)_{,j} \cr
&={1\over 3}\nabla\cdot\left({1\over 2}\nabla\cdot
\left(\vec v \nabla\cdot\vec v - \vec v \cdot \nabla \vec v \right) \vec v +
\left(\vec v \nabla\cdot\vec v - \vec v \cdot \nabla \vec v \right)\cdot\nabla
\vec v \right) \cr
&= {1\over 9}\theta^3 + 2\theta (\sigma^2 + {1\over 3}\omega^2 )
+ \sigma_{ij}\sigma_{jk}\sigma_{ki} - \sigma_{ij}\omega_i \omega_j
\;\;,&(8c)\cr}
$$
where ${\cal V}_{ij}$ is the matrix with the subdeterminants of $u_{i,j}$ 
as elements. The second equalities in (8a--c) show that all invariants can be
expressed in terms of divergences of vector fields (which has been 
used and discussed in the context of perturbation solutions -- 
see Buchert 1994).
In obtaining them, the flatness of space is used essentially.

The velocity gradient $v_{i,j}=v_{(i,j)} + v_{\lbrack i, j\rbrack}$
has, in general, $6$ independent scalar invariants:
$$
\theta \;\;,\;\; \sigma \;\;,\;\; \omega \;\;,\;\;\tau := {1\over 6}
\sigma_{ij}\sigma_{jk}\sigma_{ki} \;\;,\;\; \sigma_{ij}\omega_i \omega_j 
\;\;,\;\; \sigma_{ij}\sigma_{jk}\omega_i \omega_k \;\;\;,\eqno(8d) 
$$
and determines an invariant, orthonormal triad, the eigen--triad of the 
shear tensor; these data together with the $3$ Euler--angles
of the triad characterize the $9$ elements of $v_{i,j}$ 
invariantly at any event.

Truesdell's invariant, dimensionless vorticity measure (see Serrin 1959)
is equal to
$$
\mu := {\omega \over \sqrt{\sigma^2 + {1\over 6}\theta^2}} \;\;.
\eqno(8e)
$$
All these kinematical variables can be expressed in terms of $\vec f$
and its derivatives by means of eqs. (3).
\medskip
It is useful to define the Lagrangian (``comoving'') time--derivative
of a spatial differential form (such as $\bfo$ in equation (6e))
as the partial $t-$derivative, taken at fixed $X_i,dX_i$. (For the
intrinsic, geometrical meaning of this derivative see Appendix B.)

Then, (6e) implies
$$
\dot{\bfo} = d{\ddot f}_i \wedge df_i = d(g_i dx_i) = d{\bf g}
\;\;.\eqno(9)
$$
Therefore, we have the following {\bf kinematical Lemma}:

\noindent
Let $\vec v \lbrack \vec x,t\rbrack$ be a
(continuously differentiable) velocity field and $\vec g = {\dot{\vec
v}}$ the corresponding acceleration field. Then $\vec g$
is irrotational, $\nabla \times \vec g = \vec 0$, if and only if its
vorticity two--form $\bfo$ is conserved in the sense that
$$
{\dot\bfo} = {\bf 0} \;\;,\;{\rm i.e.}\;,\;\;{\bfo}_t = {\bfo}_{t_0}\;\;.\eqno(10)
$$
(For several equivalent formulations see Appendix B.)
\bigskip\bigskip
\bigskip
\noindent
{\rma 2.2. Dynamics of self--gravitating ``dust''}
\bigskip\noindent
So far we considered only kinematical relations which hold for any
regular
flow field $\vec f$. We now formulate the dynamical equations
for Newtonian self--gravitating flows, restricting attention
to pressureless matter (``dust'') throughout this paper.
Henceforth the variables $x_i$ are to be interpreted
as orthonormal coordinates of a dynamically non--rotating frame of
reference.

\bigskip
\noindent
{\rmb 2.2.1. Conservation of mass}
\bigskip\noindent
In the Lagrangian framework
{\it mass--conservation} states that for a regular flow
$$
\varrho(\vec X,t) =
{1\over J (\vec X,t)}\ueber{\varrho}{o}(\vec X)\;\;.\eqno(11a)
$$
The Eulerian mass density $\varrho$ can be calculated from (11a) by
using the inversion map $\vec h \lbrack\vec x,t\rbrack$:
$\varrho \lbrack \vec x,t \rbrack = \varrho
(\vec h \lbrack\vec x,t\rbrack,t)$.

Given $\ueber{\varrho}{o}(\vec X)>0$, we have shown that under the
assumptions of Subsection 2.1.1, $\varrho$ is finite and positive
for $t_0 \le t \le t_1$. If, contrary to those assumptions, $J
\rightarrow 0$, then $\varrho \rightarrow \infty$.

In terms of differential forms equation (11a) states that the
density
three--form $\varrho \;d^3 x = \varrho \;dx_1 \wedge dx_2 \wedge
dx_3$ is constant along the flow $\vec f$:
$$
\varrho \;d^3 x \;=\;\ueber{\varrho}{o} \;d^3 X \;\;,
\eqno(11b)
$$
hence ${d\over dt}(\varrho \;d^3 x)$ $= {\dot\varrho}\;d^3 x + \varrho
\;3 dv_{\lbrack 1} \wedge dx_2 \wedge dx_{3\rbrack}$ $=
({\dot\varrho}+\varrho\nabla \cdot \vec v)d^3 x = 0$, i.e.,
$$
{\dot\varrho} + \varrho \theta = 0 \;\;.\eqno(11c)
$$
\vfill\eject
\noindent
{\rmb 2.2.2. Gravitational field equations}
\bigskip\noindent
For regular flows, ``Newton's'' gravitational field equations,
generalized by a cosmological term,
$$
\nabla \times \vec g = \vec 0 \;\;\;;\;\;\;
\nabla \cdot \vec g = \Lambda - 4\pi G\varrho
\;\;,\eqno(12a,b,c,d)
$$
are, in view of equations (4), equivalent to the system of four
Lagrangian evolution equations (obtained first by Buchert \& G\"otz
1987 ($\Lambda=0$) and Buchert 1989 ($\Lambda \ne 0$)):
$$
{\cal J}({\ddot f}_j, f_j, f_k)\;=0\;\;,
\eqno(13a,b,c)
$$
$$
\left({\cal J}({\ddot f}_1, f_2, f_3)+{\cal J}({\ddot f}_2,f_3,f_1)
+{\cal J}({\ddot f}_3,f_1,f_2)\right) - \Lambda \;J\;=-4\pi G
\ueber{\varrho}{o}
\;\;.\eqno(13d)
$$
Expressed in terms of differential forms, the {\it Lagrange--Newton
system} (13) reads:
$$
d{\ddot f}_j \wedge df_j \;=\;d({\ddot f}_j df_j) \;=0\;\;,
\eqno(13a,b,c)
$$
and
$$
3 d{\ddot f}_{\lbrack 1} \wedge df_2 \wedge df_{3\rbrack} - \Lambda
\left(df_1 \wedge df_2 \wedge df_3 \right) = - 4\pi G \ueber{\varrho}{o}
\left(dX_1 \wedge dX_2 \wedge dX_3 \right)\;\;.\eqno(13d)
$$
(We keep the numbering (a,b,c) here to remind the reader that
these are in fact three equations).
Equation (13d) can also be written more compactly by using the Hodge
star operator (with respect to the metric
$\ueber{dx}{\rightarrow}^2$):
$$
*d(*{\ddot f}_j df_j) = \Lambda - 4\pi G \varrho \;\;,\eqno(13d)
$$
where $\varrho$ is given by the integral (11a).

\medskip
The {\it kinematical Lemma} stated at the end of Subsection 2.1.3
shows that in the case of ``dust'', eqs. (12a,b,c) are {\it equivalent}
to the vorticity conservation law (10) which, in this case,
acquires the status of a law of gravitational dynamics,
$d{\dot f}_i \wedge df_i = {\bfo}_{t_0}$.
In particular for
irrotational ``dust''--flows, $\bfo = {\bf 0}$, the only remaining local
law of gravity is the divergence law (12d), but the equations
$d{\dot f}_i \wedge df_i = {\bf 0}$ must not be forgotten!

\medskip
The equations (13) are invariant under constant rotations $\cal R$
and time--dependent translations $\cal T$,
$$
\vec f (\vec X,t) \mapsto {\cal R} \cdot \vec f (\vec X,t) + {\cal T}(t)
\;\;,\eqno(14a)
$$
which correspond to the transformations
$$
x^{a'} = {\cal R}^{a'}_b x^b + {\cal T}^{a'} (t)
\eqno(14b)
$$
of the Eulerian coordinates. With respect to (14b), the components of
the gravitational field strength $\vec g$ transform according to
$$
g^{a'}\lbrack x^{c'},t\rbrack = {\cal R}^{a'}_b g^b \lbrack x^c ,t
\rbrack + {\ddot{\cal T}}^{a'} (t) \;\;.
\eqno(14c)
$$
In contrast to the case of {\it isolated systems}, where one
puts $\Lambda = 0$ and restricts attention to inertial frames and
Galilean transformations (${\ddot{\cal T}}=0$), in {\it cosmology}
the assumption of large--scale homogeneity does not allow to single
out some coordinate systems as inertial ones, and the inhomogeneous
term in (14c) unavoidably occurs in transformations relating
dynamically equivalent coordinate systems (Heckmann \& Sch\"ucking
1955, 1956). Then, eq. (14c) shows that the gravitational field
strength can no longer be considered as a spatial vector field
independent of the spacetime coordinate system. We shall come back
to this well--known, but frequently disregarded fact in Subsection 3.1.1. --
The arbitrariness in the choice of ${\cal R}$ and ${\cal T}$
can be restricted or even removed by
{\it global} conditions depending on the solutions considered,
as we shall see later.

\vfill\eject
\noindent
{\rmb 2.2.3. Relations between the Eulerian and the Lagrangian
formulations}
\bigskip\noindent
The equations (13) are second--order evolution equations for the single
dynamical
field--variable $\vec f$. An evolution equation for the density is not
needed, since $\varrho$ is given explicitly by (11).
Thus, only three functions of four variables determine the
evolution of the system. In the Eulerian picture we have seven functions
of four variables,
e.g., the density, and the three components
of the velocity and the acceleration
field, obeying first--order equations.

Nevertheless, the {\it regular} solutions of the two systems
(those with regular kinematics in the sense of Subsection 2.1.1)
are in one--to--one correspondence, as follows from the preceding
considerations and has been indicated in (Buchert 1992).
More general solutions of either system exist, but in general they
are no longer equivalent to solutions of the other system;
see Remark (ii) below.

\bigskip\bigskip
\noindent
{\bf Remarks:}
\smallskip\noindent
(i) The transition Lagrange $\rightarrow$ Euler is simpler than
the converse process: in the former case, only the equations
$\vec x =\vec f (\vec X,t)$ have to be solved ``algebraically''
for $\vec X$, whereas in the other case, one has to solve the
differential equations (1) for $\vec f$.
\smallskip\noindent
(ii)
In writing the first version of
equations (13a,b,c,d) we dropped the factor
$J^{-1}$ in front of all terms. This is, of course, permitted as long
as $J \ne 0$; it holds in particular for {\it regular} solutions.
Since those {\it equations} are regular
even at singularities of the system of flow lines, i.e., where $J=0$,
and, in general, $J$ changes sign,
one may consider Lagrangian solutions which have caustics or
intersecting trajectories. One may {\it define}
$\varrho\lbrack \vec x , t \rbrack = \sum_i
{\ueber{\varrho}{o}({\vec X}_i)\over | J({\vec X}_i , t) |}$, where
the sum is performed over all values
${\vec X}_i$ such that $\vec f ({\vec X}_i , t) = \vec x$.
Such solutions, which contain ``multi--dust''
regions, are no longer equivalent to Eulerian ones.
{\it Their physical meaning and validity
requires separate considerations and is by no means
obvious}. In particular, they cannot be considered as weak limits of
Vlasov--Poisson solutions, since in the multi--stream
region particles at the same place with different velocities in general 
have different accelerations, which violates the weak principle of equivalence.
A general--relativistic theory for multi--dust spacetimes which does {\it not} suffer
from this defect, has been outlined by 
Clarke \& O'Donnell (1992). It would seem to be useful to develop a corresponding
Newtonian theory. Compare also discussions of this problem by Gurevich \&
Zybin (1995).

\vfill\eject

\noindent{\rmm 3. Newtonian Cosmology in Lagrangian Form}
\bigskip\bigskip
\noindent
{\rma 3.1. Basic concepts and equations}
\bigskip\noindent
{\rmb 3.1.1. Euclidean and toroidal cosmological models}
\bigskip\noindent
In Newton's original theory, which was designed and well--defined
for isolated systems only, as well as in standard versions of
``Newtonian'' cosmology (see, e.g., Heckmann \& Sch\"ucking 1955, 1956,
or Heckmann 1968), physical space is assumed to be ``the''
Euclidean space based on the manifold $\R^3$.
For some purposes it is useful {\it or even necessary} to model $3-$space
as closed, i.e., compact without boundary, as we shall argue in
Subsection 3.1.3. It is indeed possible to do that
without changing any of the local laws so far adopted.

Since a closed, locally Euclidean $3-$space is isometric to the
quotient of a flat torus by a finite
group of isometries\footnote*{in particular, it cannot have the topology of 
a 3--sphere, a fact
which excludes ``Newtonian'' cosmological models based on a 3--sphere} (Kobayashi \& Nomizu 1963),
we may without loss of
generality take space to be such a torus ${\Bbb T}^3$.
It is then still possible to cover space at each time by finitely
many overlapping orthonormal coordinate systems related by
transformations (14b) with ${\ddot{\cal T}} \ne 0$.

The inhomogeneous transformation law (14c) for the gravitational
field strength can be understood by reformulating Newton's theory
in covariant spacetime language as initiated by Cartan (1923, 1924)
and completed by Trautman (1966) (see also the recent 
work on Newton--Cartan cosmology by Rueede \& Straumann (1996)). 
In that reformulation the
gravitational field is represented as a symmetric, linear connection
on spacetime, as in General Relativity. It then turns out
that there exist non--rotating orthonormal local coordinates
$(t,x^{a})$ such that the only non--vanishing components of the
connection are given by $\Gamma^a_{t t}$. Moreover, the transformations
relating these coordinates are those given by (14b), and with respect to
them the $\Gamma^a_{t t}$ transform exactly like the $g^a$.
In fact, the free--fall law ${\ddot x}^a = g^a$, rewritten as the
geodesic equation ${\ddot x}^a + \Gamma^a_{t t} = 0$, shows that
we have the identity $g^a = - \Gamma^a_{t t}$, which ``explains'' the
inhomogeneous transformation law and will prove useful below.

\bigskip\noindent
{\rmb 3.1.2. Existence of solutions}
\bigskip\noindent
Neither the Euler--Newton system, nor the Lagrange--Newton system
is a differential system to which standard existence theorems apply.
The first system is mixed hyperbolic--elliptic, while the second is
an overdetermined implicit system not fitting into the standard
classification of PDE theory;
the latter may better be considered as an ordinary differential
equation for the evolution of the time--dependent displacement map.
(In this respect, the analogous equations
of General Relativity are ``simpler'' (Four\`es--Bruhat 1958).)
Nevertheless, Brauer (1992) succeeded in proving
linearization stability of the Euler--Newton system at spatially
compact (i.e. periodic) Friedmann--like
solutions and
local--in--time existence and uniqueness of solutions
which represent finite perturbations of those cosmological models,
and Brauer \etal (1994) strengthened this result in several ways.
The existence and uniqueness results established in these papers
refer to deviations from a spatially compact homogeneous background model which
has to be specified, at least partly, for all time and not just
by initial data; they do not refer to the total solution (background
+ perturbation). In fact, ``the field equations of the Newton--Cartan
theory [a 4--dimensional reformulation of ``Newton's'' theory],
unlike the Einstein equations, ``are not strong enough to determine
a solution uniquely in terms of initial data'' (Brauer \etal 1994).
For this and other reasons, work in Newtonian cosmology should be
considered as a step towards corresponding relativistic considerations.
\smallskip

Known solutions of the Lagrangian equations
include Newtonian analogs of Friedmann's and Bianchi--type
general--relativistic cosmological models.
Some exact inhomogeneous solutions have also
been found (see Subsection 3.2.3).

\bigskip\noindent
{\rmb 3.1.3. Locally isotropic cosmological models}
\bigskip\noindent
Those fluid motions which are locally isotropic in the sense that,
at any time and for each fluid particle $\cal P$, there exists a
neighbourhood on which the field of velocities relative to $\cal P$
is invariant under all rotations about $\cal P$, are characterized by
$\omega = 0$, $\sigma=0$, $\nabla \theta = 0$ and given with
our coordinate choice (1b) by
$$
\vec x = \vec f_H (\vec X,t) = a(t) \vec X
\;\;,\;\; a(t_0):=1\;\;, \eqno(15)
$$
if we conventionally put $\vec f_H (\vec 0,t) = \vec 0$.
Such a motion, a {\it Hubble flow},
solves the Euler--Newton or the Lagrange--Newton system, respectively,
if and only if Friedmann's equation holds,
$$
{{\dot a}^2 - e \over a^2} =
{8 \pi G \varrho_H + \Lambda \over 3} \; ; \; e =const.\;,
\eqno(16)
$$
which implies
$$
{\ddot a \over a} = {-4 \pi G \varrho_H + \Lambda \over 3}
\;\;,\eqno(16')
$$
where $\varrho_H = \varrho_H (t_0) a^{-3}$ denotes
the homogeneous density, and $e$, $\Lambda$ and
$\varrho_H (t_0)$ are constants.
Equation (16) holds as well in General Relativity, where the energy constant
$e$ is related to the Gaussian curvature $K_0$ at $t_0$ by
$e = - K_0 c^2$. Local isotropy implies spatial homogeneity, as is
well--known.
\medskip
Instead of considering the 3--spaces $t=const.$ of the 
locally isotropic, Friedmann--like
solutions as globally Euclidean, we may
consider the latter as closed, i.e., without loss of generality
as toroidal, as remarked above. The simplest case arises 
if we identify all those points
(particles) whose Lagrangian coordinates differ by integer multiples
of some constant length $L$ (for the general case see Brauer \etal
1994).
In order not to burden our equations
by powers of $L$, let us choose $L$ as our unit of length, i.e., put
$L=1$. All particles of such a {\it toroidal universe} change their
distances in proportion to $a(t)$, the locally Euclidean metric
is $\ueber{dx}{\rightarrow}^2 = a^2 (t)
\ueber{dX}{\rightarrow}^2$ as before, but now the
total volume of the universe is $a^3 (t)$. Note that this universe
is homogeneous and locally, but not globally isotropic.
The coordinate lines $X^a = const.$ correspond to the shortest
closed geodesics (of length $L=1$); geodesics of different directions
may be closed and longer, or not closed and of infinite length. If
we fix an orientation (handedness), the coordinate system $(X^a)$ is
now intrinsically fixed except for translations and those rotations
which map the preferred orthonormal triad onto itself. This removes
the arbitrariness of $\cal R$ in eq. (14a) except for the $9$
rotations just mentioned.

The toroidal space as a differentiable manifold cannot be covered in
a one--to--one, bicontinuous manner by a single coordinate system.
The coordinates $(X^a)$ used so far are coordinates on ${\R}^3$,
the covering space of the torus ${\Bbb T}^3$. In order to see whether
the gravitational field is well--defined on the spacetime with
toroidal space, it is inconvenient to use Eulerian coordinates
$(x^a)$ and the corresponding $g^b = - \Gamma^b_{t t}={\ddot a\over a}
x^b$; for then one would have to cover ${\Bbb T}^3$ by several
overlapping Eulerian coordinate systems and use the inhomogeneous
transformations to relate the $g^a$--components in the overlap regions.
It is easier and more elegant to transform the connection components
$\Gamma^b_{t t}$ via the geodesic equation ${\ddot x}^b - {\ddot a \over
a}x^b = 0$ to the $X^a$--coordinates. Since $x^b = a(t) X^b$, we obtain
${\ddot X}^b + 2{\dot a \over a}{\dot X}^b = 0$, for arbitrarily
moving test particles (not to be confused with the particles following
the cosmological flow). Consequently, the non--vanishing components
of the gravitational connection are $\Gamma^b_{t c} = {\dot a \over a}
\delta^b_c$. This formula shows immediately that the
connection passes from ${\R}^3$ to
${\Bbb T}^3$. In fact, instead of working
``instrinsically'' on ${\Bbb T}^3$, we may use coordinates
$(X^b)$ on ${\R}^3$, with the agreement that coordinate values $(X^a)$ 
differing by integers $(N^a)$ label the same point of ${\Bbb T}^3$,
and provided the relevant fields are
periodic. The $\Gamma^b_{t c}$ are not only periodic, but
translation and rotation invariant due to the homogeneity and local
isotropy of the model.
(This is not obvious in terms of Eulerian components.)
\medskip
\noindent
In Subsections 3.1.5 and 3.2 we shall consider inhomogeneous models
as (finite) deviations from ``Friedmann''--models on ${\Bbb T}^3$, using
``periodic'' Lagrangian coordinates $(X^a)$. The reason for using 
${\Bbb T}^3$ instead of ${\R}^3$ is as follows. We shall set up a
sequence of perturbation equations and show that {\it on ${\Bbb T}^3$
the solutions} to these equations to any order exist and {\it are uniquely
determined by initial data}, in accordance with a
non--perturbative result of Brauer \etal (1994). {\it On $\R^3$}, however,
the corresponding solutions are determined, at each order, up to
harmonic functions only, i.e., {\it there are infinitely many solutions for
the same data}.

Uniqueness can also be achieved on ${\R}^3$
by restricting the perturbations to be square--integrable.
Such perturbations, however, contradict large--scale homogeneity.
Moreover, it is usual to work with periodic perturbations, which
can conveniently be represented by (discrete) Fourier series.
In any case, {\it on} ${\Bbb T}^3$, {\it but not in general on ${\R}^3$, it is
possible to relate initial and final perturbations unambiguously}.
\smallskip
\noindent
{\bf Remark:}

\noindent
We can also discuss this problem from a statistical point of view:
If one represents the typical features of the Universe not by one
solution, but by an ensemble, one can maintain {\it statistical
homogeneity} (Bertschinger 1992) in terms of an ensemble consisting of
square--integrable members, i.e., in terms of perturbations $\vec P$ on $\R^3$
(introduced below) satisfying $\int d^3 X \; {\vec P}^2 (\vec X) < \infty$.
Plancherel's theorem asserts that then the perturbations are also square--integrable in
Fourier space, i.e., $\int d^3 k \; \ueber{|{\vec P}|}{\wedge}^2 (\vec
k) < \infty$.
Additionally, we may then choose the power spectrum of the density perturbations
to obey fall--off conditions which guarantee square--integrability 
of the whole random field. Provided that
{\it all individual members} of the statistical ensemble are
square--integrable (not merely statistical averages), we can set
limits on the exponent of a power spectrum of power law form
$\Bbb{P} \propto \vert \vec k \vert^n$: On the small--scale end
($\vert k \vert\rightarrow
\infty$) we have to require $n < -3$, and on the
large--scale end ($\vert k\vert\rightarrow 0$), $n \ge -3$
(Here we refer to the
relations (27a,b) given below and the well--known relation between
peculiar--velocity and density contrast in the linear regime).
Actually, the large--scale asymptotics can be satisfied easily,
where $n\sim +1$ according to the COBE observations, but the
small--scale asymptotics is logarithmically divergent for $n = -3$,
and the maximally allowed slope is $n\sim -3$ if the spectrum
is, e.g., truncated exponentially. The latter requirement is
at the border of what is allowed in current structure formation
scenarios.

Nevertheless, as we have shown in (Buchert \& Ehlers 1996), spatially closed
universes (i.e., those which are compact without boundary) 
are singled out as the only {\it generic models} in which the averaged variables
of inhomogeneous fields represent homogeneous solutions. Thus, the toroidal 
universe is the simplest among those Newtonian cosmologies. 

\bigskip\noindent
{\rmb 3.1.4. Average properties of general}

{\rmb $\;\;\;\;$inhomogeneous cosmological models}
\bigskip\noindent
Following Buchert \& Ehlers (1996) we discuss spatial averages of 
inhomogeneous Newtonian cosmological models by deriving 
the general expansion law which is obtained by averaging 
Raychaudhuri's equation (Raychaudhuri 1955):    
$$
\eqalignno{
{\dot\theta} &= \Lambda - 4\pi G\varrho - {1\over 3}\theta^2
+2(\omega^2 - \sigma^2 )\;\;.&(17)\cr}
$$
(Differentiation of the expansion scalar
$\theta$ with respect to the time yields
$$
\dot \theta = v_{i,i,t} + v_j v_{i,i,j} = v_{i,t,i} + (v_{i,j}v_j)_{,i}
- v_{i,j}v_{j,i} = g_{i,i} + 2 \omega^2 - 2 \sigma^2 - {1\over 3}
\theta^2 \;\;.\eqno(17')
$$
In view of (12d) we obtain (17).)
\smallskip
Equation (34) shows that if,
on one trajectory, ${1\over 2}\Lambda + \omega^2 \le 2\pi G \varrho
+ \sigma^2$ (in particular, if $\Lambda = 0$ and $\omega = 0$)
and $\theta (t') \ne 0$, then there exists an instant of time $t''$
such that ${\rm sgn}(t' - t'') = {\rm sgn}(\theta(t')),
\vert t' - t'' \vert
\le {3 \over \vert \theta (t')\vert};  {\rm lim}_{t \rightarrow t''}
\varrho (t) = {\rm lim}_{t \rightarrow t''}
\vert \theta (t)\vert = \infty$.
\medskip\noindent
Let us consider an arbitrary  ``comoving'' (Lagrangian)
volume $V(t) =:a_{\cal D}^3 (t)$ of a 
spatially compact portion ${\cal D}(t)$ of the fluid; it changes 
according to 
$$
\dot V = {d\over dt}\int_{{\cal D}(t)}d^3 x = \int_{{\cal D}(t_0)}d^3 X \;{\dot J}
= \int_{{\cal D}(t)}d^3 x \;\theta\;\;,
$$
which may be written
$$
\langle\theta\rangle_{\cal D} = {\dot V \over V} = 
3 {{\dot a}_{\cal D}\over a_{\cal D}}\;\;.
\eqno(18)
$$
Here and in the sequel, $\langle A\rangle_{\cal D}={1\over V}\int_{\cal D} 
d^3 x \,A$ denotes the spatial average of a (spatial)
tensor field $A$ on the domain ${\cal D}(t)$ occupied by the amount of fluid 
considered, and $a_{\cal D}$ is the scale factor of that domain.
\smallskip\noindent
The average of Raychaudhuri's equation may then be written (Buchert \& 
Ehlers 1996):
$$
3{{\ddot a}_{\cal D} \over a_{\cal D}} + 4\pi G {M\over a_{\cal D}^3 } -
\Lambda = {2\over 3}\left(\langle\theta^2 \rangle_{\cal D} - 
\langle\theta\rangle_{\cal D}^2 \right) +
2\langle\omega^2 - \sigma^2 \rangle_{\cal D} \;\;.\eqno(19)
$$ 
We have used the definitions (6g,h).
Equation (19) shows that the presence of inhomogeneities affects the expansion
law which only coincides with Friedmann's law (16'), $a_{\cal D}\equiv a$,  
provided shear, vorticity and 
fluctuations of the expansion scalar vanish or cancel each other, respectively.

\noindent
Introducing the averages 
$$
\Theta : = \langle\theta\rangle_{\cal D} \;\;;\;\;\Sigma_{ij} : = 
\langle\sigma_{ij}\rangle_{\cal D} 
\;\;;\;\;\Omega_{ij} := \langle\omega_{ij}\rangle_{\cal D}  \;\;,\eqno(20a,b,c)
$$
we define a linear ``background velocity field'' $\vec V$ on $\cal D$ by $V_i := H_{ij}x_j$ with
$$
V_{i,j} =\Sigma_{ij}+{1\over 3}\Theta \delta_{ij} + \Omega_{ij} = : H_{ij}\;\;.\eqno(20d)
$$
(Note that all average variables, like $a(t)$, $\Theta (t)$, $\Sigma_{ij} (t)$ and 
$\Omega_{ij} (t)$, depend on content, shape and position of the spatial domain $\cal D$.)

While the velocity fields $\vec v$ and $\vec V$ depend on the choice of a non--rotating
frame of reference ({\it cf.} eq. (14b)) and are consequently not global vector fields on 
a toroidal model, 
the peculiar velocity field, defined as  $\vec u : = \vec v - \vec V$, always {\it is} 
a global vector field. 
Splitting expansion, shear and vorticity into their (time--dependent) average parts 
and deviations thereof,
$$
\theta = \Theta + \hat{\theta} \;\;;\;\;\sigma_{ij} = \Sigma_{ij} + \hat{\sigma}_{ij}
\;\;;\;\;\omega_{ij} = \Omega_{ij} + \hat{\omega}_{ij}\;\;,\eqno(21a,b,c) 
$$ 
equation (19) can be cast into the form
$$
3{{\ddot a}_{\cal D} \over a_{\cal D}} + 4\pi G {M\over a_{\cal D}^3 } -
\Lambda = 2(\Omega^2 - \Sigma^2 ) + {2\over 3}\langle\hat{\theta}^2 \rangle_{\cal D} +
2\langle\hat{\omega}^2 - \hat{\sigma}^2 \rangle_{\cal D}\;\;.\eqno(22)
$$
(The averages $\langle\hat{\theta}\rangle_{\cal D}$, 
$\langle\hat{\sigma}_{ij}\rangle_{\cal D}$ and $\langle\hat{\omega}_{ij}\rangle_{\cal D}$
vanish by definition.)

\noindent 
Using (8b) for the peculiar--velocity 
gradient $(u_{i,j})$,
$$
{2 \over 3}\hat{\theta}^2 + 2(\hat{\omega}^2 - \hat{\sigma}^2 ) \;=\;
\nabla\cdot \lbrack \vec u (\nabla \cdot \vec u ) - (\vec u \cdot \nabla)
\vec u \rbrack \;\;,
$$
we finally arrive at the remarkably simple general expansion law:
$$
3{{\ddot a}_{\cal D} \over a_{\cal D}} + 4\pi G {M\over a_{\cal D}^3 } -
\Lambda = 2(\Omega^2 - \Sigma^2 ) + 
\langle\nabla\cdot \lbrack \vec u (\nabla \cdot \vec u ) - 
(\vec u \cdot \nabla) \vec u \rbrack \rangle_{\cal D} \;\;.\eqno(23)
$$
The last term in (23) is, via Gau{\ss}'s theorem, a surface integral over the 
boundary of $\cal D$. 
In case of a toroidal model we may choose $\cal D$ to be the whole torus. 
Thus, on the torus, we obtain the global expansion law
(in agreement with the result of Brauer \etal (1994)): 
$$
3{{\ddot a}_{\cal D} \over a_{\cal D}} + 4\pi G {M\over a_{\cal D}^3 } -
\Lambda = 2(\Omega^2 - \Sigma^2)\;\;;\;\;{\cal D} = {\Bbb T}^3 \;\;.\eqno(23')
$$
This law, combined with the linearity of the velocity field $\vec V$, can 
be used to determine all homogeneous, in general anisotropic Newtonian models
either on $\R^3$ or on ${\Bbb T}^3$, in Eulerian or Lagrangian form
(for models on $\R^3$ in Eulerian form, see Heckmann \& Sch\"ucking (1959)).

The point of this subsection was to show how these models arise by spatially 
averaging arbitrary inhomogeneous models, provided either space is 
compact or, if for ${\cal D} \rightarrow \R^3$, the last term in (23) vanishes.  

\bigskip\noindent
In the remainder of this paper we restrict ourselves to models having locally isotropic
backgrounds, i.e., where $\Sigma_{ij} = 
\Omega_{ij} = 0$; then, the average 
motion is a Hubble flow whose expansion is described by Friedmann's law (16').

\bigskip\noindent
{\rmb 3.1.5. Inhomogeneous cosmological models}

{\rmb $\;\;\;\;$as deviations from locally isotropic ones}
\bigskip\noindent
We wish to consider periodic or toroidal inhomogeneous models which 
are isotropic (and hence irrotational) {\it on average} on some large scale.
As shown in the last subsection,  
the requirement of periodicity implies that the spatially averaged density
$$
\langle\varrho\rangle>_{{\Bbb T}^3}(t): = {\int_{{\Bbb T}^3} d^3 X \; \ueber{\varrho}{o} 
\over \int_{{\Bbb T}^3} d^3 X \;
J(\vec X,t)} = {M_{tot} \over V(t)} = {M_{tot} \over a^3 (t)}
\;\;.\eqno(24)
$$
of any such model is related to $a(t)$ by Friedmann's
equation (16) with some
constants $e$, $\Lambda$, $\varrho_H (t_0)$
(which are then uniquely determined).
Thus, we can associate with any inhomogeneous model its toroidal locally
isotropic
{\it background model} defined by $\varrho_H :=\langle\varrho\rangle_{{\Bbb T}^3}$ and $a(t)$
via eqs. (15), (16), as described in Subsection 3.1.3.

To describe inhomogeneous cosmological models we 
define the deviation $\vec p$ of the displacement map
$\vec f$ of the inhomogeneous model from the
background model ${\vec f}_H$:
$$
\vec f = \vec f_H + \vec p (\vec X,t)\;\;;\;\;\vec p (\vec X,t_0):=\vec
0  \;\;.\eqno(25a,b)
$$
It is convenient to introduce periodic rescaled Eulerian coordinates
\footnote*{i.e., Lagrangian coordinates of the background flow},
$\vec q := \vec x / a(t)$ and the corresponding deformation field $\vec F$,
$\vec q = \vec F (\vec X,t); \vec F(\vec X,t_0)=\vec X$. Then,
the equations (25) read:
$$
\vec F = \vec X  + \vec P (\vec X,t)\;\;;\;\;\vec P (\vec X,t_0):=\vec
0  \;\;,\eqno(26a,b)
$$
where $\vec P = \vec p / a(t)$. $\vec P_t : {\R}^3
\rightarrow {\R}^3$ is periodic and may be interpreted as the (conformally
rescaled) displacement of
the particles of the perturbed flow relative to those of the unperturbed
flow. It
is considered the fundamental object of Lagrangian perturbation theory
hereafter.

To fix the (fictitious) {\it mean displacement} of the perturbed flow
relative to the unperturbed one (``identification gauge condition''),
we require, without loss of generality, besides (26b) for all $t$:
$$
\int_{{\Bbb T}^3}\;d^3 X \;\vec P (\vec X,t) \;=\;\vec 0
\;\;.\eqno(26c)
$$
It fixes the choice of $\cal T$ in equation (14a) and
is essential for the uniqueness of
Newtonian solutions, as we shall see later.
Note that (26c) can also be written 
$\langle\varrho / \ueber{\varrho}{o} \;\;\vec P \rangle_{{\Bbb T}^3} = 0$ so that,
if $\ueber{\varrho}{o}$ is nearly constant, $\langle\varrho \;\vec P \rangle_{{\Bbb T}^3} 
\approx 0$, a center--of--mass condition.

\smallskip
The displacement vector $\vec P$ determines the {\it
peculiar--velocity} $\vec u$ and the
{\it peculiar--acceleration} $\vec w$ by:
$$
\vec u : = \vec v - {\dot a \over a} \vec x = a \dot {\vec P} \;\;;\;\;
\ueber{\vec u}{o}=\dot {\vec P}(t_0)\;\;,
\eqno(27a)
$$
$$
\vec w := \vec g - {\ddot a \over a} \vec x = a
\ddot {\vec P} + 2 {\dot a} \dot {\vec P}\;\;;\;\;
\ueber{\vec w}{o} = \ddot {\vec P}(t_0) + 2 {\dot a}(t_0) \dot {\vec P}(t_0)
\;\;,
\eqno(27b)
$$
where $\ueber{\vec u}{o}$ and $\ueber{\vec w}{o}$ are the initial data
for peculiar--velocity
and peculiar--acceleration, respectively.
(Note that while $\vec P$, $\vec u$, $\vec w$ are {\it global} vector fields 
on ${\Bbb T}^3$,
the Hubble velocity ${\dot a \over a}\vec x$ and $\vec v$ are defined only locally with
respect to some ``origin''.)

\noindent
Below we shall use
the corresponding one--forms denoted by ${\bf U}=\ueber{u}{o}_i dX_i$
and ${\bf W}=\ueber{w}{o}_i dX_i$, and for the time--dependent
perturbation ${\bf P}=P_i dX_i$.
\bigskip\noindent
Let us now write down the equations which the displacement $\vec P$ has to obey.
Inserting (26a) into the once integrated Lagrangian evolution equations (13a,b,c)
results in
$$
d \dot {P}_i \wedge (dX_i + d{P}_i)
= a^{-2}{\ueber{\bfo}{o}}
= d(a^{-2}{\bf U})\;\;.\eqno(28a,b,c)
$$
The latter equality follows from (6e) and the fact that
the Hubble--velocity is assumed to be irrotational.
The last equation may be rewritten as
$$
d \Bigl\lbrace \dot {\bf P} + {\dot P}_i dP_i  - a^{-2} {\bf U}
\Bigr\rbrace = {\bf 0}
\;\;.\eqno(28a,b,c)
$$
Note that there is no cubic term in these equations.
\medskip\noindent
Inserting (26a) into (13d), and defining the operator ${\cal D} :=
{d^2 \over dt^2} + 2 H {d\over dt}$ and the function $b
:=3{\ddot a \over a}-\Lambda$, we obtain
$$
\eqalignno{
&b \;dX_1 \wedge dX_2 \wedge dX_3 +
({\cal D} + b) 3 dP_{\lbrack 1}\wedge dX_2 \wedge
dX_{3\rbrack}\;
+\;({\cal D} + 2b) 3 dP_{\lbrack 1}\wedge dP_2 \wedge
dX_{3\rbrack}\cr
+\;&({1\over 3}{\cal D} + b) 3 dP_{\lbrack 1}\wedge dP_2
\wedge dP_{3\rbrack} \;=\;{-4\pi G \ueber{\varrho}{o} \over a^3}
dX_1 \wedge dX_2 \wedge dX_3\;\;.&(28d')\cr}
$$
(Remember that expressions of the form $3 dA_{\lbrack 1} \wedge
dA_2 \wedge dA_{3\rbrack}$ are equal to the sum of all
cyclic permutations: $\sum_{ijk} dA_i \wedge dA_j \wedge dA_k$.)

Since this equation holds for the background, 
${\bf P} = {\bf 0}$, the terms independent of $\vec P$ cancel, and we
are left with the equation
$$
\eqalignno{
&({\cal D} + b) 3 dP_{\lbrack 1}\wedge dX_2 \wedge
dX_{3\rbrack}\;
+\;({\cal D} + 2b) 3 dP_{\lbrack 1}\wedge dP_2 \wedge
dX_{3\rbrack}\cr
+\;&({1\over 3}{\cal D} + b) 3 dP_{\lbrack 1}\wedge dP_2
\wedge dP_{3\rbrack} \;=\;{-4\pi G \delta\ueber{\varrho}{o} \over a^3}
dX_1 \wedge dX_2 \wedge dX_3\;\;,&(28d)\cr}
$$
where $\delta\ueber{\varrho}{o}=\ueber{\varrho}{o}-\ueber{\varrho_H}{o}$
is the (finite) initial deviation from the homogeneous density
$\varrho_H = \ueber{\varrho_H}{o}a^{-3}$; 
$\int_{{\Bbb T}^3} d^3 X \delta\ueber{\varrho}{o} = 0$.

In what follows we shall use the Hodge star operator
with respect to the metric
$\ueber{dX}{\rightarrow}^2$.
Therefore, we indicate it
with a big star ($\bs$) to avoid confusion with the Hodge star operator
used in previous equations. (The following identities
are useful: $\bs d^3 X = 1$, $(\bs)^2 =1$,
$d\bs d\bs = \bs d\bs d = \Delta_0$.)

Operating with $\bs$ on (28d) and using $4\pi G
\delta\ueber{\varrho}{o}=\bs d\bs W$, gives
$$
\bs d\Bigl\lbrace({\cal D}  + b) {\bf \bs P} \;
+\;({\cal D} + 2b) 3 P_{\lbrack 1}
\wedge dP_2 \wedge dX_{3\rbrack}\;+\;({1\over 3}{\cal D} + b)
3 P_{\lbrack 1}\wedge dP_2 \wedge
dP_{3\rbrack} - a^{-3}{\bf \bs W}\Bigr\rbrace\;=0\;\;.\eqno(28d)
$$
Here, the linear term is purely longitudinal.

\noindent
The equations (28a,b,c,d) with the initial conditions (26b) govern 
inhomogeneous models.
\smallskip
In more familiar vector notation the equations (28a,b,c,d) have the
form:
$$
{d\over dt} (\nabla_0 \times \vec P) = \ueber{\cal F}{\rightarrow}
(\partial {\dot P}_i , \partial P_j ) + a^{-2} \nabla_0 \times
\ueber{\vec u}{o}
\;\;;
$$
$$
({\cal D} +b) (\nabla_0 \cdot \vec P ) = {\cal G}
(P_i, \partial P_j , {\dot P}_i , {\ddot P}_i )
+ a^{-3} \nabla_0 \cdot \ueber{\vec w}{o} \;\;.
$$
The r.h.s.'s contain no terms linear in $\vec P$ or its derivatives, and
they contain no derivatives with respect to $t$ or $X_i$ of higher order
than on the l.h.s.$\;$. Therefore, these equations lend themselves to
solution by iteration. For that purpose, the condensed differential
form notation is more convenient than vector notation, however.

\bigskip\bigskip\noindent
{\rma 3.2. Lagrangian perturbation theory}
\bigskip\noindent
{\rmb 3.2.1. The perturbation scheme}
\bigskip\noindent
Since we have only one dynamical object in the problem (the one--form
${\bf P}$), a Lagrangian perturbation scheme on
Friedmann-Lema\^\i tre backgrounds can be set up by inserting
into eqs. (28) for ${\bf P}$ a formal power series,
$$
{\bf P}
= \sum_{m = 1}^{\infty} \varepsilon^{m} {\bf P}^{(m)}
\;\;,\eqno(29)
$$
to obtain a sequence of equations for the ${\bf
P}^{(m)}$ at order $m$.
We thus obtain the following system of $4m$
equations:

\noindent
For $m=1$ we have
$$
d {{\dot{\bf P}}^{(1)}} = d (a^{-2}{\bf U}^{(1)} ) \;\;;\eqno(30a,b,c;
m=1)
$$
$$
d \bs\Bigl\lbrace\Bigl\lbrack
{\cal D} + b
\Bigr\rbrack {\bf P}^{(1)}
\Bigr\rbrace \;=\;d(a^{-3} \bs {\bf W}^{(1)}) \;\;\;.\eqno(30d;m=1)
$$
For $m>1$ we have
$$
d \Bigl\lbrace{{\dot{\bf P}}^{(m)}} \Bigr\rbrace
= d {\bf T}^{(m)} \;\;;\eqno(30a,b,c;m>1)
$$
$$
d \bs\Bigl\lbrace\Bigl\lbrack {\cal D} + b
\Bigr\rbrack {{\bf P}^{(m)}} \Bigr\rbrace
=d\bs{\bf S}^{(m)} \;\;\;.
\eqno(30d;m>1)
$$
The $2m$ source terms (one--forms) ${\bf S}^{(m)}$
and ${\bf T}^{(m)}$
can be read off eqs. (28). They depend on ${\bf P}^{(\ell)};\;\ell
< m$:
$$
{\bf T}^{(m)}
= - \sum_{\ell =1}^{m-1} {{\dot P}^{(\ell)}}_i
d {P^{(m-\ell)}}_i  + a^{-2}{\bf U}^{(m)}
\;\;,\eqno(31a;m>1)
$$
$$
\bs {\bf S}^{(m)}=-\sum_{\ell=1}^{m-1} ({\cal D} + 2b)
3{P^{(\ell)}}_{\lbrack 1}
d{P^{(m-\ell)}}_2 \wedge dX_{3\rbrack}\;-\;
$$
$$
\sum_{\scriptstyle \ell+p+q=m \atop\scriptstyle
1\le \ell,p,q \le m-2}({1\over 3}{\cal D} + b)
3{P^{(\ell)}}_{\lbrack 1} d{P^{(p)}}_2
\wedge d{P^{(q)}}_{3\rbrack} + a^{-3}\bs{\bf W}^{(m)}\;\;.\eqno(31b;m>1)
$$
Starting at the third order, the source terms contain products of
perturbation solutions of different orders,
(compare Buchert 1994, eqs. (4)).

\bigskip\bigskip
\noindent
{\rmb 3.2.2. General solution scheme}
\bigskip\noindent
To solve the equations (30) with the source terms (31), we
decompose the ${\bf P}^{(m)}$'s as well as the initial
values ${\bf U}$ and ${\bf W}$ {\it non--locally}
into their longitudinal and transverse
parts (see Appendix C),
$$
{\bf P}^{(m)} = {{\bf P}^{(m)}}^L + {{\bf P}^{(m)}}^T
\;\;,\eqno(32a)
$$
$$
{\bf U}^{(m)} = {{\bf U}^{(m)}}^L + {{\bf U}^{(m)}}^T
\;\;,\eqno(32b)
$$
$$
{\bf W}^{(m)} = {{\bf W}^{(m)}}^L \;\;\;\;\;\;\;\;\;\;\;\;\;
\;\;,\eqno(32c)
$$
taking into account that {\it the harmonic parts vanish} because of the
gauge condition (26c) and eqs. (27), and remembering
that $d{\bf W} = {\bf 0}$.

We prescribe, without loss of generality,
that the initial density perturbation
and thus ${\bf W}$ be of first order,
$$
\delta\ueber{\varrho}{o} = \delta\ueber{\varrho}{o}^{(1)}
=:  \ueber{\varrho}{o}_H \ueber{\delta}{o}
\;\;;\;\;{\bf W}^{(1)} = {\bf W}
\;\;,\eqno(33a,b)
$$
where $\delta\ueber{\varrho}{o}$ denotes the initial density
perturbation, and
$\ueber{\delta}{o}$ the initial (conventional) density contrast.

Equation (26b) requires, for all $m$,
$$
{\bf P}^{(m)} (\vec X,t_0):={\bf 0} \;\;.\eqno(33c)
$$
Finally we require, also without loss of generality,
$$
{\dot{\bf P}} (\vec X,t_0) = {\dot{\bf P}}^{(1)} (\vec X,t_0)
={\bf U} (\vec X) \;\;.\eqno(33d)
$$
\bigskip\noindent
The {\it unique solutions} of the perturbation equations having these
initial data are obtained as follows.

\noindent
Equations $(30a,b,c;m=1)$ say that
$$
{\bf A}:={\dot{\bf P}}^{(1)^T} - a^{-2}{\bf U}^{(1)^T}
$$
is both closed, $d{\bf A} = {\bf 0}$, and co--exact, hence it
vanishes (see Appendix C); therefore
$$
{{\bf P}^{(1)}}^T (\vec X,t) = {\bf U}^T (\vec X)
\int_{t_0}^{t} {dt' \over a^2 (t')}\;\;.\eqno(34a,b,c)
$$
Eq. $(30d;m=1)$ similarly implies
$$
({\cal D} + b) {{\bf P}^{(1)}}^L (\vec X,t)
= a^{-3}{\bf W} (\vec X)\;\;.\eqno(34d)
$$
The solution to this ordinary differential equation obeying the
initial conditions (33) is uniquely determined by the data ${\bf W}
(\vec X)$ and ${\bf U}^L (\vec X)$.
\medskip\noindent
For $m>1$ we obtain from (30a,b,c):
$$
{{\bf P}^{(m)}}^T (\vec X,t)
= \int_{t_0}^{t} dt'\;{{\bf T}^{(m)}}^T (\vec X,t')\;\;;\eqno(35a,b,c)
$$
and from (30d):
$$
({\cal D} + b) {{\bf P}^{(m)}}^L (\vec X,t) = {{\bf S}^{(m)}}^L
\;\;.\eqno(35d)
$$
The solutions to eqs. (35) are uniquely determined by their sources
(31), since they are required to have vanishing initial values.

\bigskip\noindent
{\bf Remarks:}
\smallskip\noindent
(i) The solutions at any order $m$
are well--defined and unique on $\R \times {\Bbb T}^3$ as long as
the background is free of singularities.
In general they will develop ``multi--dust'' regions.
\smallskip\noindent
(ii) The solutions at any order $m$
separate with respect to Lagrangian coordinates $\vec X$ and time $t$;
${\bf P}^{(m)} (\vec X,t) = \sum_{\alpha} A_{\alpha}^{(m)} (\vec X)
B_{\alpha}^{(m)} (t)$.
This property follows from the
structure of the perturbation scheme, since
the first--order solutions separate and, at each step,
only linear ordinary differential equations with respect to $t$ have
to be solved. The time--dependent coefficients are determined solely
by the background, while the $\vec X$--dependent factors depend on
the initial data.
\smallskip\noindent
(iii) The first--order solution depends locally on the data
${\bf U}$ and ${\bf W}$ in the sense that the factors
$A_{\alpha}^{(1)} (\vec X)$ at some value $\vec X$ depend only on
${\bf U}$ and ${\bf W}$ at the same $\vec X$. On the other hand,
${\bf W}$ depends non--locally, via a solution of Poisson's equation,
on $\ueber{\delta}{o}$. Each further step involves the determination of
${{\bf T}^{(m)}}^T$ and ${{\bf S}^{(m)}}^L$ from ${\bf T}^{(m)}$ and
${\bf S}^{(m)}$, respectively, which again requires to solve
Poisson equations. Thus, the $\vec X$--dependent factors
in ${\bf P}^{(m)}$ depend non--locally on the data ${\bf U}$ and
${\bf W}$ for $m>1$.
The trajectory of each ``dust particle'' at any order of
approximation depends {\it globally} on the initial data, even at times
close to the initial time, just as in Newtonian dynamics of systems of
finitely many particles. This is in contrast to General Relativity,
where the evolved fields at some spacetime point depend
only on the initial data within the causal past of that point.
(For GR ``dust'' solutions this has first been shown by
Four\`es--Bruhat 1958.)

\smallskip\noindent
(iv) Since all relevant functions are defined on ${{\Bbb T}^3}$, they
can be represented by discrete
Fourier series. Since the sources for the
higher--order terms are products of lower--order ones, the higher--order
terms will change on smaller spatial scales than the lower--order ones,
and their time--dependent factors will contain (positive and negative)
powers of those of the first--order solution which generates the
higher--order ones.
\smallskip\noindent
(v) If the perturbation scheme is applied to fields on $\R^3$
rather than on ${\Bbb T}^3$, at each step a harmonic contribution to
${\bf P}^{(m)}$ has to be chosen arbitrarily. (This is due, of course,
to the form of eqs. (12)). Then, there are infinitely many
perturbative solutions for given initial data; hence,
{\it it makes no sense to ask which fields evolve from which data}.
\smallskip\noindent
(vi) The equations (34) suggest that it is convenient to introduce a
new time--variable $T$ (taken to be dimensionless):
$$
dT:={1\over t_0} {dt\over a^2 (t)}\;\;.\eqno(36a)
$$
This variable has been very useful for the purpose of finding solutions
for ``non--flat'' backgrounds (see: Shandarin (1980), Buchert
(1989, Appendix A), Bouchet \etal 1995, Catelan 1995).
With this time--variable solutions of (16) for $\Lambda =0$
have the simple form:
$$
a(T) = {K_0 + T_0^2 \over K_0 + T^2} \;\;.\eqno(36b)
$$
Also the time--dependent operator in front of the longitudinal part
simplifies ($\Lambda \ne 0$ here):
$$
t_0^2 ({\cal D} + b)
= {d^2\over dT^2} - 4\pi G \ueber{\rho_H}{o} a
\;\;.\eqno(36c)
$$
(Compare: Buchert (1989, Appendix A) for the Lagrangian equations
as well as all relevant cosmological
variables and parameters expressed in terms of $T$).

\bigskip\bigskip

\noindent
{\rmb 3.2.3. Explicit solutions}
\medskip\noindent
(Not in chronological order of their derivation.)
\medskip\noindent
Known solutions comprise the general first--order solution
(Buchert 1992) for an ``Einstein--de Sitter'' background, which 
includes rotational flows and the
``Zel'dovich Approximation'' (Zel'dovich 1970, 1973) as the special case
${\bf U}^T = {\bf 0}$, ${\bf U}^L = {\bf W}t_0$.

For irrotational flows the solution for all backgrounds with $\Lambda =
0$ can be found in (Buchert 1989) including generalizations of
Zel'dovich's approximation obtained by Shandarin (1980).

For most of the background solutions including
a cosmological constant, closed--form expressions are given in
(Bildhauer \etal 1992), where a general procedure to obtain
the ``Zel'dovich Approximation'' for all backgrounds is outlined.

Interestingly, for restricted initial data,
the first--order solutions turn out to be exact three--dimensional 
solutions (Buchert
1989) including the general plane--symmetric solution
given earlier by Zentsova \& Chernin (1980).
These solutions contain caustics.
(For related exact solutions see Buchert \& G\"otz (1987), Barrow \& G\"otz (1989) and
Silbergleit (1995).)

At second order all irrotational solutions on an Einstein--de Sitter
background are known for initial data
which admit a functional dependence of initial peculiar--velocity
and peculiar--gravitational potentials (Buchert \& Ehlers 1993).
A subclass of these solutions for the special case
${\bf U}^T = {\bf 0}$, ${\bf U}^L = {\bf W}t_0$
is discussed in Buchert (1993). For the same initial data
the third--order solution on an Einstein--de Sitter background is given
by Buchert (1994), the fourth--order solution by Vanselow (1995); 
see Sahni \& Coles (1995) and Buchert (1996a,b) for reviews.

Lagrangian perturbation solutions and their applications have also been 
derived and applied
by Bouchet \& collaborators (for a review see Bouchet \etal (1995), 
where references to solutions
with ``non--parabolic'' cosmological backgrounds at second 
(Bouchet \etal 1992) and third 
order for the leading time coefficient (the particular solutions)
can be found). Moutarde \etal (1991) gave a third--order approximation 
on an Einstein--de Sitter background for special symmetric initial data.
For these data a (slightly different) solution has been derived from the
generic solution by Buchert \etal (1996).
The general irrotational second--order solution for
``non--parabolic'' cosmological backgrounds with zero cosmological constant
has been derived by Vanselow (1995).
Also Munshi \etal (1994) discuss the leading terms of the
third--order solution
of Buchert (1994), and Catelan (1995) derives and discusses
the third--order solution for ``non--parabolic'' backgrounds.
\bigskip
The main difference between most of these works and our approach is that
we consistently work within the Lagrangian framework, i.e., we
express all equations in terms of the single dynamical field $\vec f$
{\it before} solving them. Hence, we avoid mixing of Lagrangian and
Eulerian representations. The only perturbed field
is $\vec f$ in Lagrangian space; all Eulerian fields are calculated
therefrom. The velocity field is determined perturbatively, the
corresponding mass and the vorticity is exactly conserved
in our perturbation solutions.

\medskip
The fundamental question whether these perturbation solutions
converge to or, at least, approximate exact solutions remains open.

\bigskip\bigskip\bigskip
\noindent
{\rma Acknowledgements:} TB is supported by the ``Sonderforschungsbereich 
375 f\"ur
Astro--Teilchenphysik der Deutschen Forschungsgemeinschaft''. 
He would like to thank 
the Albert--Einstein--Institut in Potsdam, where parts of this work have been 
written, for generous support and hospitality.

\vfill\eject

\centerline{\rmm APPENDIX A}
\bigskip\bigskip\noindent
Under the assumptions stated at the
beginnning of Subsection 2.1.1, the map $f_t: {\Bbb D}^3 \rightarrow
{\Bbb D}^3$;
${\Bbb D} \in \lbrace \R, {\Bbb T} \rbrace$; $t$ fixed; ($t_0 \le t
\le t_1$) is a diffeomorphism.
We first show that $f_t$ is injective, and then that it is surjective.
Since $f_t$ is a local diffeomorphism because of $J>0$, this
establishes the claim.
\medskip\noindent
{\it Injectivity} follows immediately from the fact that different
integral--curves of a vector field
are disjoint.
\medskip\noindent
To establish {\it surjectivity} we notice the following:

Since
$f_{t}$ is a local diffeomorphism, the image $f_t({\Bbb D}^3)$ is open.
It is also closed; for let ${\vec x}_i = {\vec f}_t ({\vec X}_i)$ be
a sequence of images which converges to ${\vec x}_0$, ${\vec x}_i
\rightarrow {\vec x}_0$. Then, the set $\lbrace {\vec X}_i \rbrace$
is bounded since $\lbrace {\vec x}_i \rbrace$ is, and distances can
change during $\lbrack t_0,t \rbrack$ at most by $2V \vert t-t_0 \vert$.
Therefore, a subsequence of $\lbrace {\vec X}_i \rbrace$ converges to
some point ${\vec X}_0$. Continuity of $f_t$ then implies that
${\vec x}_0 = {\vec f}_t ({\vec X}_0) \in f_t ({\Bbb D}^3)$. Thus,
$f_t ({\Bbb D}^3)$ is both open and closed in ${\Bbb D}^3$,
hence equal to ${\Bbb D}^3$.

\bigskip\bigskip

\centerline{\rmm APPENDIX B}
\bigskip\bigskip\noindent
We here give an invariant meaning to the
``time--differentiation'' of
differential forms which was used in the main text (the reader
may consult standard textbooks on differential forms, e.g., Schutz 1980),
and we collect different versions of the vorticity conservation law
$\dot{\bfo}={\bf 0}$.
\medskip
\noindent{\sl Lie--derivative}
\medskip\noindent
We defined the operator $\;\dot{}\;$ on spatial differential forms 
as partial differentiation
with respect to $t$ for fixed $\vec X$. In Newtonian spacetime
$\R \times \R^3$ or $\R \times {\Bbb T}^3$, a velocity field
$\vec v \lbrack \vec x,t \rbrack$ determines a world velocity field,
$$
{d\over dt} = {\partial \over \partial t} + v_i {\partial \over
\partial x_i} \;\;.\eqno(B.1)
$$
If we use Lagrangian coordinates ($\vec X,t$) on spacetime, the
vector field ${d\over dt}$ has components ($\vec 0,1$). Therefore,
in these coordinates, Lie--differentiation with respect to
${d\over dt}$ amounts to partial differentiation with respect to $t$.
This shows that
$$
{\bf L}_{d\over dt} {\bf A} = \dot{\bf A} \eqno(B.2)
$$
for all ``spatial'' differential forms, i.e.,
differential forms not containing $dt$, and gives the invariant
meaning of $\;\dot{}\;$. This time--derivative commutes with
spatial exterior differentiation, $d$.
\medskip
We now list some equivalent versions of the vorticity conservation
law
$$
\dot{\bfo} = {\bf 0} \;\;,\eqno(B.3a)
$$
since different versions appear in the literature and are useful
for different purposes (for all these relations it is necessary
that the force is conservative, i.e. the gravitational field
strength $\vec g$ is irrotational).
\smallskip
The vector form of (B.3a) reads:
$$
\dot{\vec\omega} = {\vec\omega} \cdot \nabla \vec v
- {\vec\omega} \nabla \cdot \vec v
\;\;.\eqno(B.3b)
$$
We can integrate $\vec \omega$ along the integral--curves
$\vec f$ to obtain Cauchy's integral
(see, e.g., Serrin 1959, Buchert 1992),
$$
{\vec \omega} = (\ueber{\vec \omega}{o}\cdot \nabla_0 \vec f) \,J^{-1}
\;\;.\eqno(B.3c)
$$
Equation (B.3c) shows that the vorticity blows
up at points of (formally) infinite density ($J=0$)
for
generic initial data (see Buchert 1992 for a proof). This implies that
caustics
are associated with strong vortex flows in their vicinity
(see also the detailed discussion by Barrow \& Saich 1993).

In terms of kinematical variables, the vorticity law reads:
$$
{\dot\omega}_i = -{2\over 3}\theta \omega_i + \sigma_{ij}\omega_j
\;\;.\eqno(B3.d)
$$

\bigskip\bigskip

\centerline{\rmm APPENDIX C}
\bigskip\bigskip\noindent
In order to make this paper self--contained and to fix our notation
we here collect some well--known facts about decompositions of
vector fields on $\R^3$ and ${\Bbb T}^3$, respectively, both
furnished with the standard flat (Lagrangian) metric
$\ueber{dX}{\rightarrow}^2$.

On $\R^3$, any smooth vector field $\vec P$ can be decomposed into
a gradient (longitudinal) part and a curl (transverse) part,
$$
\vec P = {\vec P}^L + {\vec P}^T = \nabla_0 U + \nabla_0 \times \vec A
\;\;,\;\;\nabla_0 \cdot \vec A = 0\;\;.\eqno(C.1)
$$
Such a decomposition always exists, whether or not $\vec P$ falls off
at infinity; but it is not unique: if $\vec H$ is a harmonic field,
i.e., a field satisfying $\nabla_0 \cdot \vec H = 0$ and
$\nabla_0 \times \vec H = \vec 0$, then
$$
\vec P = (\nabla_0 U + \vec H) + (\nabla_0 \times \vec A - \vec H )
$$
gives another representation of the type (C.1), since $\vec H = \nabla_0
\psi = \nabla_0 \times \vec B$, and in this way all such representations
are obtained. If $\vec P$ as well as the parts ${\vec P}^L$ and
${\vec P}^T$ are required to be square integrable ($\in {\cal L}^2$),
i.e., $\int d^3 X {\vec P}^2 < \infty$, the decomposition (C.1)
is unique; square integrable harmonic fields do not exist on $\R^3$
(Dodziuk 1979).
Then one can speak of {\it the} longitudinal, or {\it the} transverse
part of $\vec P$, respectively.

On ${\Bbb T}^3$, one has a unique decomposition:
$$
\vec P =  \nabla_0 U + \nabla_0 \times \vec A \;+\; \vec H
\;\;,\eqno(C.2)
$$
where the harmonic part $\vec H$ is constant on ${\Bbb T}^3$ (see the
remark below) and given by:
$$
\vec H =
\int_{{\Bbb T}^3} d^3 X \;\;
\vec P \;\;.\eqno(C.3)
$$
The potentials $U$ and $\vec A$ can also be fixed uniquely by
requiring:
$$
\int_{{\Bbb T}^3} d^3 X \;\;U \;=\; 0
\;\;,\;\;
\int_{{\Bbb T}^3} d^3 X \;\;
\vec A \;=\; \vec 0\;\;,\;\;
\nabla_0 \cdot \vec A \;=\; 0 \;\;.\eqno(C.4)
$$
Note that, on ${\Bbb T}^3$,
being longitudinal means not only that
$\nabla_0 \times \vec P = \vec 0$,
but in addition that $\int d^3 X \; \vec P \;=\;\vec 0$. Similarly,
transversality requires $\nabla_0 \cdot \vec P = 0$ and vanishing
average.

It is convenient to re--express these facts in the language of
differential forms rather than that of vector fields. Writing
${\bf P}= P_i dX_i$ for the one--form (covector) associated with $\vec
P$, the form--analogs are:
$$
{\bf P} = {\bf P}^L + {\bf P}^T + {\bf P}^H \;=\;
d U + \bs d {\bf A} + {\bf H}\;\;,\eqno(C.2')
$$
where $\bf A$ and $\bf H$ are one--forms, the longitudinal part
is an {\it exact} form, the transverse part a {\it co--exact} form,
and the harmonic part a {\it harmonic} form, which is determined by
$$
\int_{{\Bbb T}^3} d^3 X \;\;
{\bf P} = {\bf H} \;\;,\eqno(C.3')
$$
and one may impose
$$
\int_{{\Bbb T}^3} d^3 X \;\;U = 0 \;\;,\;\;
\int_{{\Bbb T}^3} d^3 X \;\;{\bf A} = {\bf 0}\;\;,\;\;
d \bs{\bf A} = 0 \;\;,\eqno(C.4')
$$
where in all equations
$\bs$ denotes the Hodge star operator with respect to the
metric $\ueber{dX}{\rightarrow}^2$.

The integration of the perturbation equations in Subsection 3.1.4
is based on the following two facts: If a co--exact form ${\bf P}^T$ is
closed, $d{\bf P}^T = {\bf 0}$, it is the zero--form, ${\bf P}^T =
{\bf 0}$. If an exact form ${\bf P}^L$ is co--closed, $d\bs {\bf P}^L =
{\bf 0}$, it is the zero--form, ${\bf P}^L = {\bf 0}$. These facts
follow from the foregoing statements and equations.

We also recall that Poisson's equation,
$$
\Delta_0 U = 4\pi G \varrho \;\;,\eqno(C.5)
$$
is soluble on ${\Bbb T}^3$ if and only if $\int_{{\Bbb T}^3} d^3 X \;
\varrho =
0$. The solution is then unique except for an additive constant which
may be fixed by demanding:
$$
\int_{{\Bbb T}^3} d^3 X \;\;U = 0 \;\;.\eqno(C.6)
$$
For proofs see, e.g., Warner (1971).

\bigskip\noindent
{\bf Remark:}
The only {\it harmonic}
vector-fields ${\vec H}$ are the constant ones.
To see this, we recall the vector--identity
$$
\Delta_0 \vec H = \nabla_0 \times (\nabla_0 \times \vec H) -
\nabla_0 (\nabla_0 \cdot \vec H) \;\;.\eqno(C.7)
$$
It shows that a harmonic vector field obeys Laplace's equation.
Then, its components $H_i \;(i=1,2,3)$ are harmonic functions.
For each component, we can apply Green's formula,
$$
\int_{{\Bbb T}^3} \;H_i \Delta_0 H_i
= \int_{{\Bbb T}^3} \;H_i \nabla_0 (\nabla_0 H_i)
= \int_{{\Bbb T}^3} \;\lbrace \nabla_0 (H_i \nabla_0 H_i) -
(\nabla_0 H_i)^2 \rbrace
$$
$$
= \int_{\partial {{\Bbb T}^3}} \; H_i {\partial H_i \over \partial
n} - \int_{{\Bbb T}^3} \;(\nabla_0 H_i)^2 \;\;.\eqno(C.8)
$$
Since the scalars $H_i$ are harmonic, the
left--hand--side of the identity (C.8) vanishes. Since
the torus ${\Bbb T}^3$ has no boundary, we finally conclude
$$
\int_{{\Bbb T}^3} \;(\nabla_0 H_i)^2 \;=\;0\;\;, \eqno(C.9)
$$
or, $\nabla_0 H_i = 0$. Hence, $H_i = const.$.

\vfill\eject

\centerline{\rmm References}
\bigskip\bigskip
\ref
Barrow J.D., G\"otz G. (1989): {\it Class. Quant. Grav.} {\bf 6}, 1253.
\ref
Barrow J.D., Saich P. (1993): {\it Class. Quant. Grav.} {\bf 10}, 79.
\ref
Bertschinger E. (1992): in: {\it Lecture Notes in Physics} {\bf 408}, p.65, 
Springer.
\ref
Bertschinger E., Hamilton A.J.S. (1994): {\it Ap.J.} {\bf 435}, 1.
\ref
Bertschinger E., Jain B. (1994): {\it Ap.J.} {\bf 431}, 486.
\ref
Bertschinger E. (1996): in {\sl Cosmology and Large Scale Structure} 
Proc. Les Houches XV Summer School, Elsevier Science Publishers B.V., 
in press.
\ref
Bildhauer S., Buchert T., Kasai M. (1992): {\it Astron. Astrophys.} {\bf
263}, 23.
\ref
Bouchet F.R., Juszkiewicz R., Colombi S., Pellat R. (1992):
{\it Ap.J. Lett.} {\bf 394}, L5.
\ref
Bouchet F.R., Colombi S., Hivon E., Juszkiewicz R. (1995):
{\it Astron. Astrophys.} {\bf 296}, 575.
\ref
Bouchet F.R. (1996): in: {\it Proc. IOP `Enrico Fermi'}, Course CXXXII 
(Dark Matter in the Universe), Varenna 1995, eds.:
S. Bonometto, J. Primack, A. Provenzale, in press.
\ref
Brauer U. (1992): {\it J. Math. Phys.} {\bf 33}, 1224.
\ref
Brauer U., Rendall A., Reula O. (1994): {\it Class. Quant. Grav.}
{\bf 11}, 2283.
\ref
Bruni M., Matarrese S., Pantano O. (1995): {\it Ap.J.} {\bf 445}, 958.
\ref
Buchert T., G\"otz G. (1987): {\it J. Math. Phys.} {\bf 28}, 2714.
\ref
Buchert T. (1989): {\it Astron. Astrophys.} {\bf 223}, 9.
\ref
Buchert T. (1992): {\it M.N.R.A.S.} {\bf 254}, 729.
\ref
Buchert T. (1993): {\it Astron. Astrophys.} {\bf 267}, L51.
\ref
Buchert T., Ehlers J. (1993): {\it M.N.R.A.S.} {\bf 264}, 375.
\ref
Buchert T. (1994): {\it M.N.R.A.S.} {\bf 267}, 811.
\ref
Buchert T., Melott A.L., Wei{\ss} A.G. (1994): {\it Astron. Astrophys.}
{\bf 288}, 349.
\ref
Buchert T. (1996a): in: {\it Proc. IOP `Enrico Fermi'}, Course CXXXII 
(Dark Matter in the Universe), Varenna 1995, eds.:
S. Bonometto, J. Primack, A. Provenzale, in press.
\ref
Buchert T. (1996b): {\it Phys. Rep.}, submitted.
\ref
Buchert T., Ehlers J. (1996): {\it Astron. Astrophys.}, in press.
\ref
Buchert T., Karakatsanis G., Klaffl R., Schiller P. (1996): 
{\it Astron. Astrophys.}, in press.
\ref
Cartan E. (1923): {\it Ann. Sci. Nom. Sup.} {\bf 40}, 325 (in French).
\ref
Cartan E. (1924): {\it Ann. Sci. Nom. Sup.} {\bf 41}, 1 (in French).
\ref
Catelan P. (1995): {\it M.N.R.A.S.} {\bf 276}, 115.
\ref
Clarke C.J.S., O'Donnell N. (1992): {\it Rend. Sem. Math. Univ. Pol.
Torino} {\bf 50}, 39.
\ref
Coles P., Melott A.L., Shandarin S.F. (1993): {\it M.N.R.A.S.}
{\bf 260}, 765.
\ref
Croudace K., Parry J., Salopek D., Stewart J. (1994): {\bf 423}, 22.
\ref
Dodziuk J. (1979): {\it Proc. Am. Math. Soc.} {\bf 77}, 395.
\ref
Ehlers J. (1961): {\it Akad. Wiss. Lit. Mainz, Abh. Math.--Nat. Klasse
11}, p.793 (in German); translated (1993): {\it G.R.G.} {\bf 25}, 1225.
\ref
Ehlers J., Buchert T. (1996): {\sl in preparation}
\ref
Ellis G.F.R. (1971): in: {\it General Relativity and Cosmology}, ed. by
R. Sachs, N.Y.: Academic press.
\ref
Ellis G.F.R., Dunsby P.K.S. (1996): {\it Ap.J.}, in press.
\ref
Four\`es--Bruhat Y. (1958): {\it Bull. Soc. Math. France} {\bf 86},
155 (in French).
\ref
Gramann M. (1993): {\it Ap.J.} {\bf 405}, L47.
\ref
Gurevich A.V., Zybin K.P. (1995): {\it Sov. Phys. Uspekhi} {\bf 38}, 687.
\ref
Heckmann O. (1968): {\it Theorien der Kosmologie}, 2. Auflage, Springer
(in German).
\ref
Heckmann O., Sch\"ucking E. (1955): {\it Zeitschrift f\"ur Astrophysik}
{\bf 38}, 95 (in German).
\ref
Heckmann O., Sch\"ucking E. (1956): {\it Zeitschrift f\"ur Astrophysik}
{\bf 40}, 81 (in German).
\ref
Heckmann O., Sch\"ucking E. (1959): {\it Encyclopedia of Physics} {\bf 53},
489, Springer (in German). 
\ref
Kasai M. (1992): {\it Phys. Rev.} {\bf D47}, 3214.
\ref
Kasai M. (1995): {\it Phys. Rev.} {\bf D52}, 5605.
\ref
Kobayashi S., Nomizu K. (1963): {\it Foundations of Differential
Geometry}, Interscience, New York.
\ref
Kofman L., Pogosyan D. (1995): {\it Ap.J.} {\bf 442}, 30.
\ref
Lachi\`eze--Rey M. (1993): {\it Ap.J.} {\bf 408}, 403.
\ref
Lesame W.M., Ellis G.F.R., Dunsby P.K.S. (1996): {\it Ap.J.}, submitted.
\ref
Matarrese S., Pantano O., Saez D. (1993): {\it Phys. Rev.} {\bf D47},
1311.
\ref
Matarrese S., Pantano O., Saez D. (1994): {\it M.N.R.A.S.} {\bf 271}, 513.
\ref
Matarrese S. (1996): in: {\it Proc. IOP `Enrico Fermi'}, Course CXXXII 
(Dark Matter in the Universe), Varenna 1995, eds.:
S. Bonometto, J. Primack, A. Provenzale, in press.
\ref
Matarrese S., Terranova D. (1996): {\it M.N.R.A.S.}, in press.
\ref
Melott A.L., Pellmann T.F., Shandarin S.F. (1994): {\it M.N.R.A.S.}
{\bf 269}, 626.
\ref
Melott A.L., Buchert T., Wei{\ss} A.G. (1995): {\it Astron. Astrophys.}
{\bf 294}, 345.
\ref
Moutarde F., Alimi J.-M., Bouchet F.R., Pellat R., Ramani A.
(1991): {\it Ap.J.} {\bf 382}, 377.
\ref
Munshi D., Sahni V., Starobinsky A.A. (1994): {\it Ap.J.} {\bf 436}, 517.
\ref
Peebles P.J.E. (1980): {\it The Large-scale Structure of the Universe},
Princeton Univ. Press.
\ref
Peebles P.J.E. (1993): {\it Principles of Physical Cosmology},
Princeton Univ. Press.
\ref
Raychaudhuri A. (1955): {\it Phys. Rev.} {\bf 98}, 1123.
\ref
Rueede C., Straumann N. (1996): {\it Helv. Phys. Acta}, in press
(preprint: gr--qc/9604054). 
\ref
Russ H., Morita M., Kasai M., B\"orner G. (1996): {\it Phys. Rev.} {\bf D53}, 
6881.
\ref
Sahni V., Coles P. (1995): {\it Phys. Rep.} {\bf 262}, 1.
\ref
Salopek D.S., Stewart J.M., Croudace K.M. (1994): {\it M.N.R.A.S.} {\bf 271},
1005.
\ref
Schutz B.F. (1980): {\it Geometrical methods of mathematical physics},
Cambridge University Press.
\ref
Serrin J. (1959): in {\sl Encyclopedia of Physics} {\bf VIII.1}, Springer. 
\ref
Shandarin S.F. (1980): {\it Astrophysics} {\bf 16}, 439.
\ref
Silbergleit A. (1995): {\it J. Math. Phys.} {\bf 36}, 847.
\ref
Stuart J.T., Tabor M. (1990): (eds) {\it The Lagrangian picture of fluid motion}
{\it Phil. Trans. R. Soc. Lond. A} {\bf 333}, 261-400.
\ref
Trautman A. (1966): in: {\it Perspectives in Geometry and Relativity},
ed. by B. Hoffmann, Indiana Univ. Press, Bloomington, pp.413-425.
\ref
Tr\"umper M. (1965): {\it J. Math. Phys.} {\bf 6}, 584.
\ref
Vanselow M. (1995): Diploma Thesis, Ludwig--Maximilians--Universit\"at
M\"unchen (in German).
\ref
Warner F.W. (1971): {\it Foundations of Differentiable Manifolds
and Lie Groups}, Scott Foresman, Glenvier, III.
\ref
Wei{\ss} A.G., Gottl\"ober S., Buchert T. (1996): {\it M.N.R.A.S.} {\bf 278},
953.
\ref
Zel'dovich Ya.B. (1970): {\it Astron. Astrophys.} {\bf 5}, 84.
\ref
Zel'dovich Ya.B. (1973): {\it Astrophysics} {\bf 6}, 164.
\ref
Zentsova A.S., Chernin A.D. (1980): {\it Astrophysics} {\bf 16}, 108.

\vfill\eject

\bye